\documentclass[a4paper,11pt]{article}
\pdfoutput=1

\usepackage{jcappub}

\usepackage[T1]{fontenc} 

\usepackage[english]{babel}
\usepackage[utf8]{inputenc}

\usepackage{color}
\usepackage{csquotes}

\usepackage{graphicx}

\usepackage[sc]{mathpazo}
\usepackage{upgreek}
\usepackage{units}
\usepackage{siunitx}

\usepackage[percent]{overpic}

\usepackage{amsmath,amssymb}

\usepackage{times}

\usepackage{xspace}

\def\ugeo{$f_\mathrm{geo}$\xspace}
\def\uce{$f_\mathrm{ce}$\xspace}
\def\vB{\vec{v} \times \vec{B}}
\def\vvB{\vec{v} \times  (\vec{v} \times \vec{B})}
\def\uv{f_{\vec{v}}}
\def\uvB{f_{\vec{v} \times \vec{B}}}
\def\uvvB{f_{\vec{v} \times  (\vec{v} \times \vec{B})}}
\def\EvB{E_{\vec{v} \times \vec{B}}}
\def\EvvB{E_{\vec{v} \times  (\vec{v} \times \vec{B})}}
\def\Egeo{E_\mathrm{geo}}
\def\Ece{E_\mathrm{ce}}
\def\ugeo{f_\mathrm{geo}}
\def\uce{f_\mathrm{ce}}
\def\dx{\,\mathrm{d}x}
\def\dr{\,\mathrm{d}r}
\def\dphi{\,\mathrm{d}\phi}

\def\xmax{X$_\mathrm{max}$\xspace}
\def\xmaxrad{X$_\mathrm{max}^\mathrm{rad}$\xspace}
\def\dxmax{$D_\mathrm{X_\mathrm{max}}$\xspace}
\def\dxmaxm{D_\mathrm{X_\mathrm{max}}}

\def\rhoxmax{$\rho_{X_\mathrm{max}}$\xspace}

\def\sina{$\sin \upalpha$\xspace}
\def\eradgeo{$E_\mathrm{rad}^\mathrm{geo}$\xspace}
\def\eradce{$E_\mathrm{rad}^\mathrm{ce}$\xspace}

\def\srdxmax{$S_\mathrm{RD}^\rho$\xspace}

\title{Simulation of Radiation Energy Release in Air Showers}

\author[a,1]{Christian Glaser,\note{Corresponding author.}}
\author[a]{Martin Erdmann,}
\author[b, c]{J\"org R. H\"orandel,}
\author[d]{Tim Huege}
\author[b]{and Johannes Schulz}

\affiliation[a]{RWTH Aachen University, III. Physikalisches Institut A, Aachen, Germany}
\affiliation[b]{IMAPP, Radboud University Nijmegen, Nijmegen, Netherlands}
\affiliation[c]{Nikhef, Science Park, Amsterdam, Netherlands}
\affiliation[d]{Karlsruhe Institute of Technology, Institut für Kernphysik, Karlsruhe, Germany}

\emailAdd{glaser@physik.rwth-aachen.de}

\abstract{
A simulation study of the energy released by extensive air showers in the form of MHz radiation is performed using the CoREAS simulation code. We develop an efficient method to extract this radiation energy from air-shower simulations. We determine the longitudinal profile of the radiation energy release and compare it to the longitudinal profile of the energy deposit by the electromagnetic component of the air shower. 
We find that the radiation energy corrected for the geometric dependence of the geomagnetic emission scales quadratically with the energy in the electromagnetic component of the air shower with a second-order dependence on the atmospheric density at the position of the maximum shower development $X_\mathrm{max}$. In a measurement where $X_\mathrm{max}$ is not accessible, this second order dependence can be approximated using the zenith angle of the incoming direction of the air shower with only a minor loss in accuracy. Our method results in an intrinsic uncertainty of 4\% in the determination of the energy in the electromagnetic air-shower component, which is well below current experimental uncertainties.}

\dedicated{doi:10.1088/1475-7516/2016/09/024}

\begin{document}
\maketitle
\flushbottom

\section{Introduction}

Ultra-high energy cosmic rays (UHECRs) produce extensive air showers when they interact in the Earth's atmosphere. Traditionally, they are measured by detecting the particles of the air shower that reach the ground, by telescopes that observe the isotropic fluorescence light emitted by molecules that have been excited by the shower particles \cite{Auger2014, TA2008} or by non-imagining air-Cherenkov telescopes that measure the incoherent Cherenkov light produced by the shower particles \cite{Tunka133_2014}. The fluorescence method gives a direct measurement of the shower development in the atmosphere. 
Accordingly, the shower development and the energy deposit resulting in the emission of fluorescence light have been studied extensively \cite{RisseHeck2004}.

Another independent method to observe air showers is the detection of broadband megahertz (MHz) radio emission which has become an active field of research in recent years \cite{LOPES2005, Codalema2006, ICRC2015JSchulz, TunkaRex2015, LOFARICRC2015, Huege2016}. 
Although processes producing radio emission are well-understood by now, a detailed study of  
how much radiation is emitted at which stage of the shower development has not been presented yet. This longitudinal shower development with respect to the radio emission will be studied in this work.

Two emission processes have been identified. The dominant geomagnetic emission arises from the deflection of electrons and positrons in the shower front at the Earth's magnetic field and is polarized along the direction of the Lorentz force ($\propto \vec{v} \times \vec{B}$) that is acting on the charged particles \cite{KahnLerche1966, LOPES2005, CodalemaGeoMag, RAuger2012}. The field strength of the emission scales with the absolute value of the geomagnetic field $\vec{B}$ and the sine of the angle $\upalpha$ between the shower direction $\vec{v}$ and the geomagnetic field. Muons are also deflected in principle, but due to their much lower charge/mass ratio they do not contribute significantly to the radio emission \cite{Huege2016}. 
The second, subdominant emission arises from a time-varying negative charge-excess in the shower front which is polarized radially with respect to the axis of the air shower \cite{Askaryan1962, Hough1970, Prescott1971, AERAPolarization, LofarPolarization2014}.

The exact amount of radio emission seen by an observer depends strongly on coherence effects \cite{Allan1971, Scholten2012} and on the position of the observer relative to the air shower. However, the radio emission arises from the acceleration of individual charged particles and is described by classical electrodynamics. Hence, radio emission by shower particles can be calculated via first principles from the well-understood electromagnetic part of an air shower. Such calculations are implemented in air shower simulation programs \cite{ZHAires2012, CoREAS2013}. In this work we use the CoREAS code \cite{CoREAS2013} which is the radio extension of the CORSIKA \cite{Corsika} software to simulate air showers. 

The total amount of radio emission emitted by an air shower can be quantified using the concept of radiation energy \cite{ICRC2015CGlaser, AERAEnergyPRD, AERAEnergyPRL}, which is the energy emitted by the air shower in the form of radio waves. In most experimental setups, the ``golden'' frequency band between \unit[30 - 80]{MHz} is used. Below \unit[30]{MHz} atmospheric noise and short-wave band transmitters make measurements unfeasible and above \unit[80]{MHz} the FM band interferes. Also, owing to coherence effects, the radio emission from air showers is strongest below \unit[100]{MHz}. Hence, in our analysis we will concentrate on this frequency band.

In an experiment, the radiation energy can be determined by interpolating and integrating the measured energy fluence on the ground \cite{ICRC2015CGlaser, AERAEnergyPRD, AERAEnergyPRL}. As soon as the air shower has emitted all radio emission, the radiation energy measured at increasing atmospheric depths remains constant because the atmosphere is essentially transparent for radio emission. In particular, the radiation energy is independent of the signal distribution on the ground that changes drastically with incoming directions of the air shower or the altitude of the observation.

In the first part of this article we will determine the ``longitudinal profile of the radiation energy release'', i.e., how much radiation energy is emitted at what stage of the shower development. 
In the second part we will determine the fundamental relation between the radiation energy and the energy in the electromagnetic part of the air shower and study second order dependences.

\section{Model}
In this section we develop a mathematical model to extract the radiation energy from air-shower simulations in an efficient manner and describe the setup of the simulations.

The three-dimensional electric field vector $\vec{E}$ of the cosmic-ray-induced radio signal can be reduced to two dimensions, because the component in the direction of propagation can be assumed to be zero with great accuracy, as an electromagnetic wave is only polarized in the plane perpendicular to its direction of propagation. This plane is usually referred to as the shower plane. An advantageous choice to align the axes in the shower plane is to align one axis with the $\vB$ direction, i.e., the direction perpendicular to the shower axis $\vec{v}$ and the geomagnetic field $\vec{B}$. The other axis is then aligned with the $\vvB$ direction to get an orthogonal coordinate system. 

The energy fluence $f$, i.e., the energy deposit per area, is the time-integral over the Poynting flux of the radio pulse and is calculated using the formula
\begin{equation}
 f = \uvB + \uvvB
\end{equation}
with the two components
\begin{eqnarray}
 \uvB(\vec{r}) =& \varepsilon_0 c \Delta t \, &\sum\limits_i \EvB^2(\vec{r}, t_i) \label{eq:energyfluence1} \\
 \uvvB(\vec{r}) =& \varepsilon_0 c \Delta t \, &\sum\limits_i \EvvB^2(\vec{r}, t_i) \, ,
 \label{eq:energyfluence2}
\end{eqnarray}
where $\varepsilon_0$ is the vacuum permittivity, c is the speed of light in vacuum and $\Delta t$ is the sampling interval of the electric field $\vec{E}(\vec{r}, t)$ which depends on the position $\vec{r}$ and time $t$ and is broken down into its components $\EvB$ and $\EvvB$.

The components $\uvB$ and $\uvvB$ can be further divided into a part originating from the geomagnetic and a part originating from the charge-excess emission processes. The geomagnetic component is polarized in the direction of the Lorentz force that is acting on the shower particles, i.e., in the $\vB$ direction, and is thus only present in the $\EvB$ component of the electric field \cite{Scholten200894}. The contribution of the charge-excess emission to the electric field is radially polarized towards the shower axis. Therefore, its contribution to $\EvB$ and $\EvvB$ depends on the detector position relative to the shower axis \cite{Vries2010a} which can be expressed by the angle $\phi$ as defined in Fig.~\ref{fig:starpattern}. The electric field components can hence be written as
\begin{eqnarray}
 \EvB(\vec{r}, t) =& E_\mathrm{geo}(\vec{r}, t) + \cos \phi \, E_\mathrm{ce}(\vec{r}, t) \\
 \EvvB(\vec{r}, t) =& \sin \phi \, E_\mathrm{ce}(\vec{r}, t)  \, ,
\end{eqnarray}
where $E_\mathrm{geo}$ ($E_\mathrm{ce}$) is the modulus of the electric field resulting from the geomagnetic (charge-excess) emission.

We define $\ugeo$ and $\uce$ as the energy fluence calculated from $E_\mathrm{geo}$ and $E_\mathrm{ce}$. From the definition of the energy fluence (Eq.~\eqref{eq:energyfluence1}+\eqref{eq:energyfluence2}) it follows that 
\begin{equation}
\label{eq:uvvB}
 \uvvB(\vec{r}) = \sin^2 \phi \, \uce(\vec{r}) \,
\end{equation}
and after some lines of calculations it follows that 
\begin{equation}
\label{eq:uvB}
 \uvB(\vec{r}) = \left(\sqrt{\ugeo(\vec{r}) } + \cos \phi \, \sqrt{\uce(\vec{r})}\right)^2 
\end{equation}
if the geomagnetic and charge-excess components are ``in phase'', i.e., $\Egeo(\vec{r}, t) = k(\vec{r}) \, \Ece(\vec{r}, t)$ for some $k \in \mathbb{R}$. Within the precision necessary for this work, this assumption is indeed valid. This approximation results in bias of 1\% resulting in an overestimation of $\uvB(\vec{r})$ of 1\%, and is discussed in appendix \ref{sec:approx1}.

\subsection{Calculation of Radiation Energy}
The energy fluence $f$ changes with the position in the shower plane which we denote in polar coordinates $(r, \phi)$ relative to the shower axis. The radiation energy, i.e., the complete energy contained in the radio signal at a given height above ground, is calculated with the formula
\begin{equation}
\label{eq:rad}
 E_\mathrm{rad} = \int\limits_0^{2\pi} \dphi \int\limits_0^{\infty} \dr \, r \, f(r,\phi) \, .
 \end{equation}
Exploiting the radial symmetry of the geomagnetic and charge-excess lateral distribution functions (LDFs), Eq.~\eqref{eq:rad} can be rewritten as
 \begin{eqnarray}
 E_\mathrm{rad} = &\int\limits_0^{2\pi} \dphi \int\limits_0^{\infty} \dr \, r \left[ \left(\sqrt{\ugeo(r)} + \cos \phi \, \sqrt{\uce(r)}\right)^2 + \sin^2 \phi \, \uce(r) \right] \\
 = &\int\limits_0^{2\pi} \dphi \int\limits_0^{\infty} \dr \, r \, \left(\ugeo(r) +  \uce(r) + 2 \cos \phi \sqrt{\ugeo(r) \, \uce(r)}\right) \\
 = &2 \pi \, \int\limits_0^{\infty} \dr \, r \, \left(\ugeo(r) +  \uce(r)\right) \, \label{eq:geocedec} \\
 = &2 \pi \, \int\limits_0^{\infty} \dr \, r \, \left(\uvB(r, \phi = 90^\circ) +  \uvvB(r, \phi = 90^\circ)\right) \, \\
 \label{eq:rad2} 
 = &2 \pi \, \int\limits_0^{\infty} \dr \, r \, f(r, \phi = 90^\circ)
 \label{eq:numint}
\end{eqnarray}
The validity of assuming radial symmetry of the geomagnetic and charge-excess LDFs is shown in appendix \ref{sec:approx2} and additional effects of the air refractivity on the symmetry of the footprint are discussed in Sec.~\ref{sec:erad_uncertainty}.

\begin{figure}[bt]
\centering
 \includegraphics[width=0.6\textwidth]{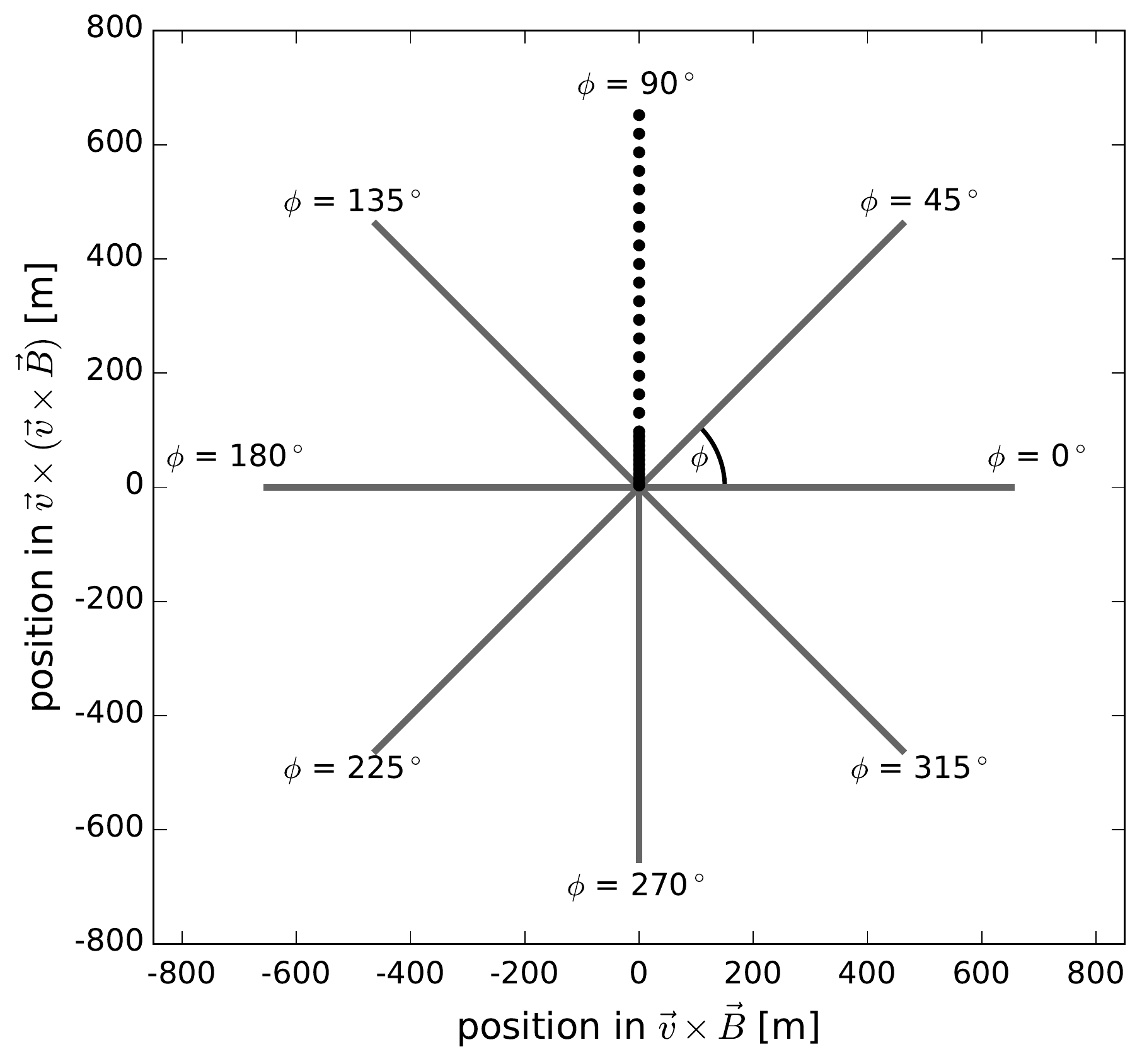}
 \caption{Sketch of the simulated observer positions. Each circle represents a simulated observer position. The observer spacing is denser near the shower axis to correctly sample the larger variations in signal strength near the shower axis.}
 \label{fig:starpattern}
\end{figure}

This result has a direct impact on how the radiation energy can be determined in Monte Carlo simulations of extensive air showers and their radio emission. Instead of simulating the radio emission at numerous positions to sample the full two-dimensional emission pattern, it is sufficient to simulate the radio signal at positions along the positive $\vvB$ axis, i.e., for $\phi = 90^\circ$. This significantly reduces the simulation time compared to the star-shaped pattern \cite{LofarXmaxMethod} (cf. Fig.~\ref{fig:starpattern}). 

We note that the component of the electric field parallel to the shower direction is not exactly zero as the radio shower front shows a slight deviation from a plane wave \cite{LOPESWavefront2014, LOFARWavefront2015}. Hence, also $\uv$ is not exactly zero but is significantly smaller than $\uvB$ and $\uvvB$. However, to exclude any potential additional uncertainty we determine $f$ of Eq.~\eqref{eq:numint} as $f = \uvB + \uvvB + \uv$.

Another benefit of this method is that along the $\vvB$ axis the geomagnetic and charge-excess contributions to the energy fluence separate ($\uvB = \ugeo$ and $\uvvB = \uce$) and can consequently be analyzed separately. In addition, the energy fluence at any position in the shower plane can be calculated from the geomagnetic and charge-excess LDFs according to Eq.~\eqref{eq:uvvB} and \eqref{eq:uvB}.  This approximation is similar to the approach presented in \cite{Alvarez-Muniz2014a}, but does away with the necessity of simulating the air shower a second time without a geomagnetic field to disentangle the geomagnetic and charge-excess components.

\subsection{Simulation Setup}
\label{sec:simulation_setup}
We use the CORSIKA 7.4005 code \cite{Corsika} to simulate extensive air showers with hadronic interaction models QGSJetII.04 \cite{QGSJet} and FLUKA 2011.2c \cite{fluka}, and the CoREAS \cite{CoREAS2013} extension to calculate the radio emission. Thinning is applied at a level of $10^{-6}$ with optimized weight limitation \cite{Kobal2001}. 

We use the model of the US standard atmosphere and set the geomagnetic field to an inclination of -35.7$^\circ$ with an absolute field strength of \unit[0.243]{Gauss}. This corresponds to the geomagnetic field at the AERA detector \cite{ICRC2015JSchulz} of the Pierre Auger Observatory \cite{Auger2014}. The influence of a different geomagnetic field as present at the LOFAR \cite{LOFARICRC2015} or the Tunka \cite{TunkaRex2015} detector will be discussed in Sec.~\ref{sec:bfield}. The refractivity of the atmosphere is set to $n - 1 = 2.92 \times 10^{-4}$ at sea level and is scaled proportionally to higher altitudes by means of air density. 

We simulate the radio emission at 30 observer positions along the positive $\vvB$ axis. The observer spacing is denser near the shower axis to improve sampling in this region with larger variations in signal strength (see Fig.~\ref{fig:starpattern})\footnote{The first station is placed at 0.5\% of the maximum distance $d_\mathrm{max}$, the next 12 stations are placed equidistantly until 15\% of $d_\mathrm{max}$ and the remaining 17 stations are again placed equidistantly until $d_\mathrm{max}$.}. According to Eq.~\eqref{eq:numint} we calculate the radiation energy by numerical integration of the energy fluence using the composite trapezoidal rule. Using a toy Monte Carlo we checked that the uncertainty due to the numerical integration is well below 1\%.

As the footprint of the radio emission becomes larger with increasing distance to the emission region we dynamically adjust the maximum distance of a simulated observer position to the shower axis\footnote{The maximum distance of an antenna to the shower axis range from {\SI{100}{m}} at an observation height that corresponds to a slant depth of {\SI{100}{g/cm^2}} to almost {\SI{3}{km}} at an observation height that corresponds to a slant depth of {\SI{5800}{g/cm^2}}.}.
All observers are placed in the shower plane, i.e., depending on the incoming direction of the air shower and their distance from the shower axis, the observer positions have a different height above ground. In the following, we refer to the height of the observer at the shower axis as the height above ground $h$ or, alternatively, by specifying the atmospheric depth $X$.

\subsection{Simulated Data Set}

The simulated data set comprises 592 air showers. The primary energy is distributed randomly between \unit[10$^{17}$]{eV} and \unit[10$^{19}$]{eV} according to a uniform distribution of the logarithm of the energy. The zenith angle $\theta$ is distributed uniformly between 0$^\circ$ and 80$^\circ$ and the azimuth angle is distributed uniformly between 0$^\circ$ and 360$^\circ$. For each combination of primary energy and shower direction, one simulation with a proton primary and one simulation with an iron primary is performed. 

In air-shower experiments, the zenith angle distribution typically follows a $\sin \theta \cos \theta$ distribution. The term $\sin \theta$ arises from the increasing solid angle for larger zenith angles and $\cos \theta$ arises because the projected area becomes smaller and radio detectors are typically triggered by particle detectors. We therefore reweight our data set in all fits and histograms to match such a zenith angle distribution. 

An important quantity for the analysis presented in the following sections is the sine of the angle $\upalpha$ between the shower axis and the magnetic-field axis. For the geomagnetic field at the Pierre-Auger site and a zenith angle distribution following $\sin \theta \cos \theta$, the $\sin \upalpha$ distribution peaks at large values of $\sin \upalpha$. A fraction of 45\% of the events has a value of $\sin \upalpha$ larger than $0.9$ and more than 97\% of the events have a value of \sina larger than $0.2$. Also for other configurations of the geomagnetic field, the number of events with low values of \sina is small. If we set the geomagnetic field to the configuration at the LOFAR \cite{LOFARICRC2015} or the Tunka \cite{TunkaRex2015} detector, still 96\% of the events have a value of \sina larger than $0.2$. In addition, the radiation energy of the dominant geomagnetic emission component is proportional to $\sin^2\alpha$. Therefore, the number of events with a small value of \sina that can be measured in an experiment is even smaller than the above estimation, which is based solely on geometry. Hence, the practical relevance of air showers with small \sina is generally small.

\section{Longitudinal Profile of Radiation Energy Release}

In this section it is determined how much radiation energy is released at what stage of the shower development. In contrast to the energy deposit of shower particles, the radiation energy cannot be accessed for different atmospheric depths in the standard output of the CoREAS simulation. As the radio pulse observed at a specific point in space is the result of constructive and destructive interference of all charged shower particles it is different at each position. Hence, it needs to be calculated separately for each observer position so that the computing time scales linearly with the number of observers. 

\begin{figure}[ptb]
\def\dx{22}
\def\dy{70}
 \centering
 \vspace{20px}
 \begin{overpic}[width=0.49\textwidth]{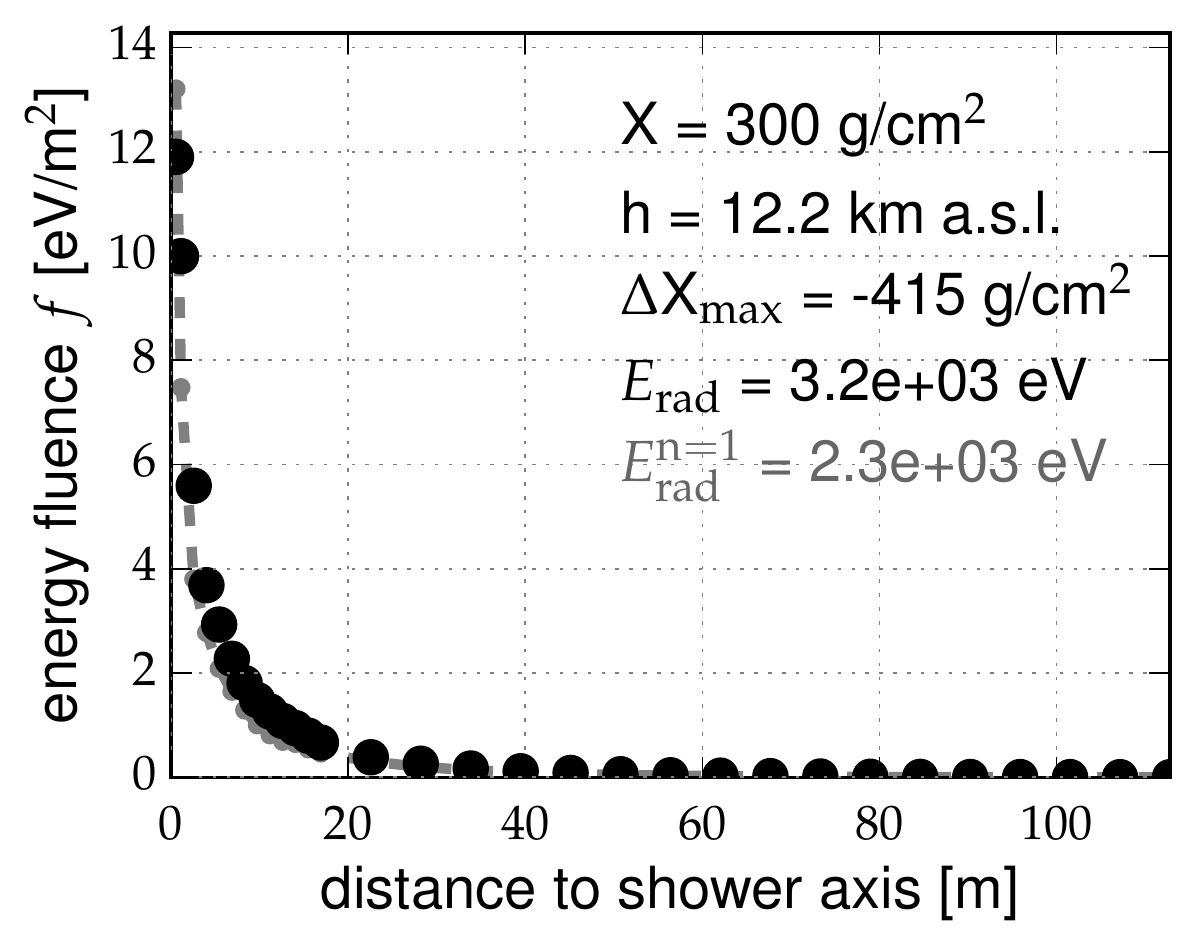}
 \put(\dx, \dy){\Large a)}
 \end{overpic}
 \begin{overpic}[width=0.49\textwidth]{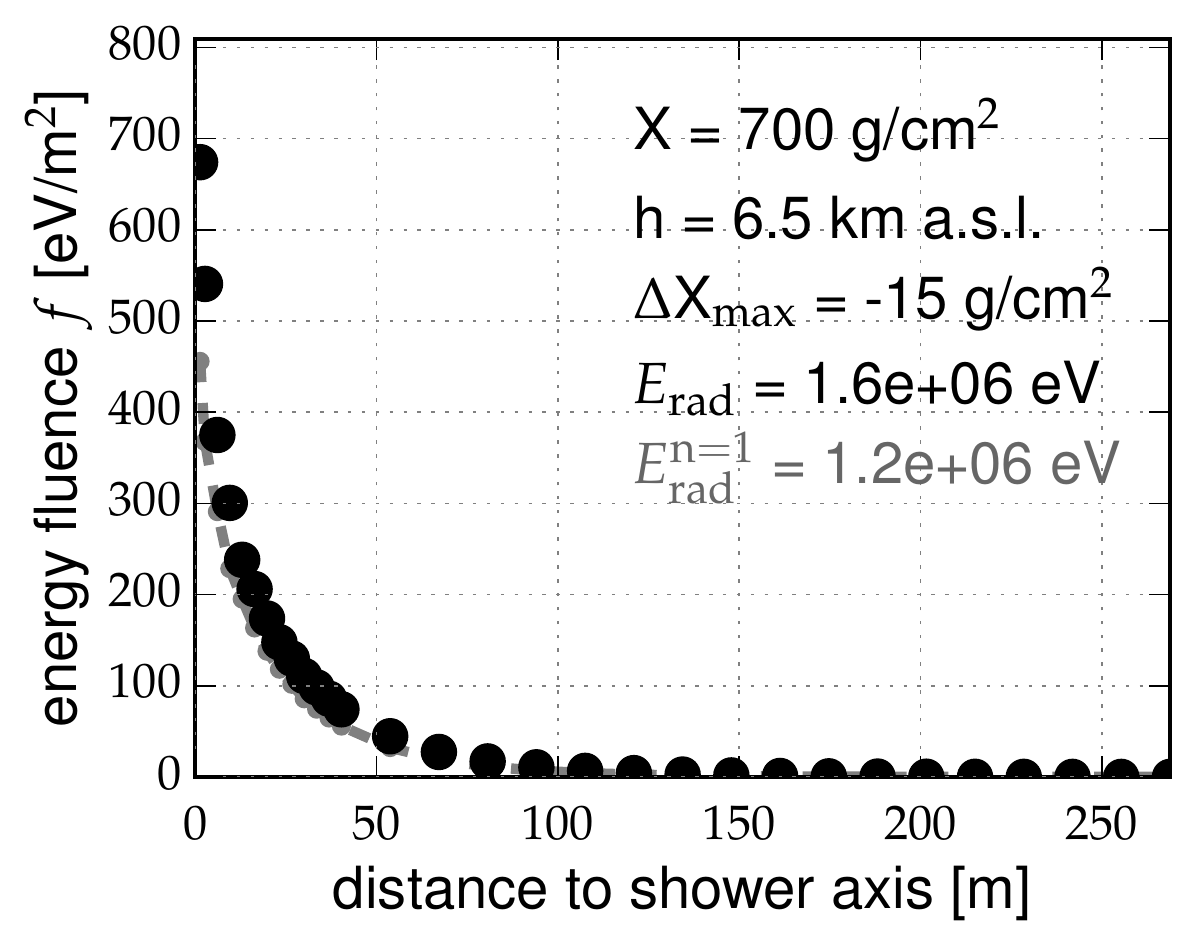}
 \put(\dx, \dy){\Large b)}
 \end{overpic}
 \begin{overpic}[width=0.49\textwidth]{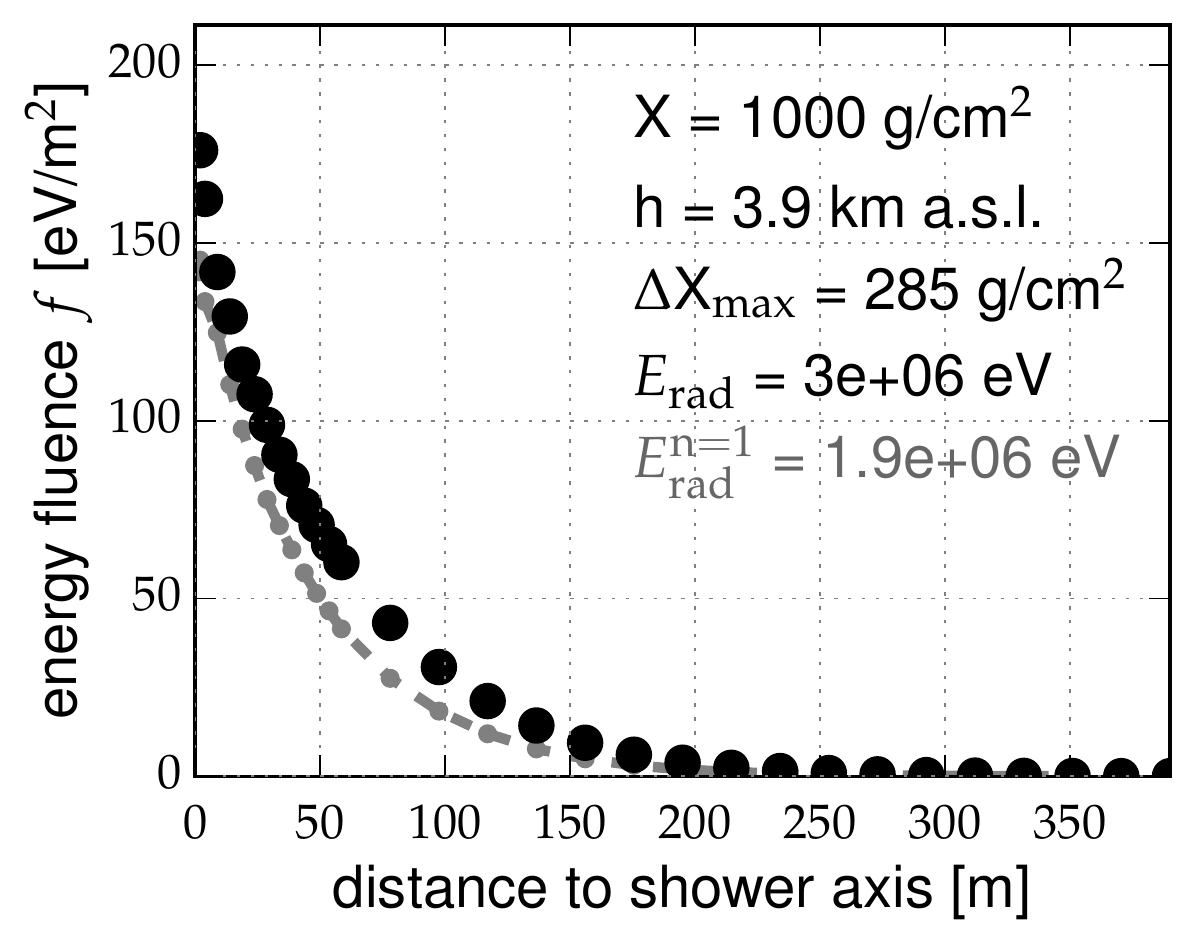}
 \put(\dx, \dy){\Large c)}
 \end{overpic}
 \begin{overpic}[width=0.49\textwidth]{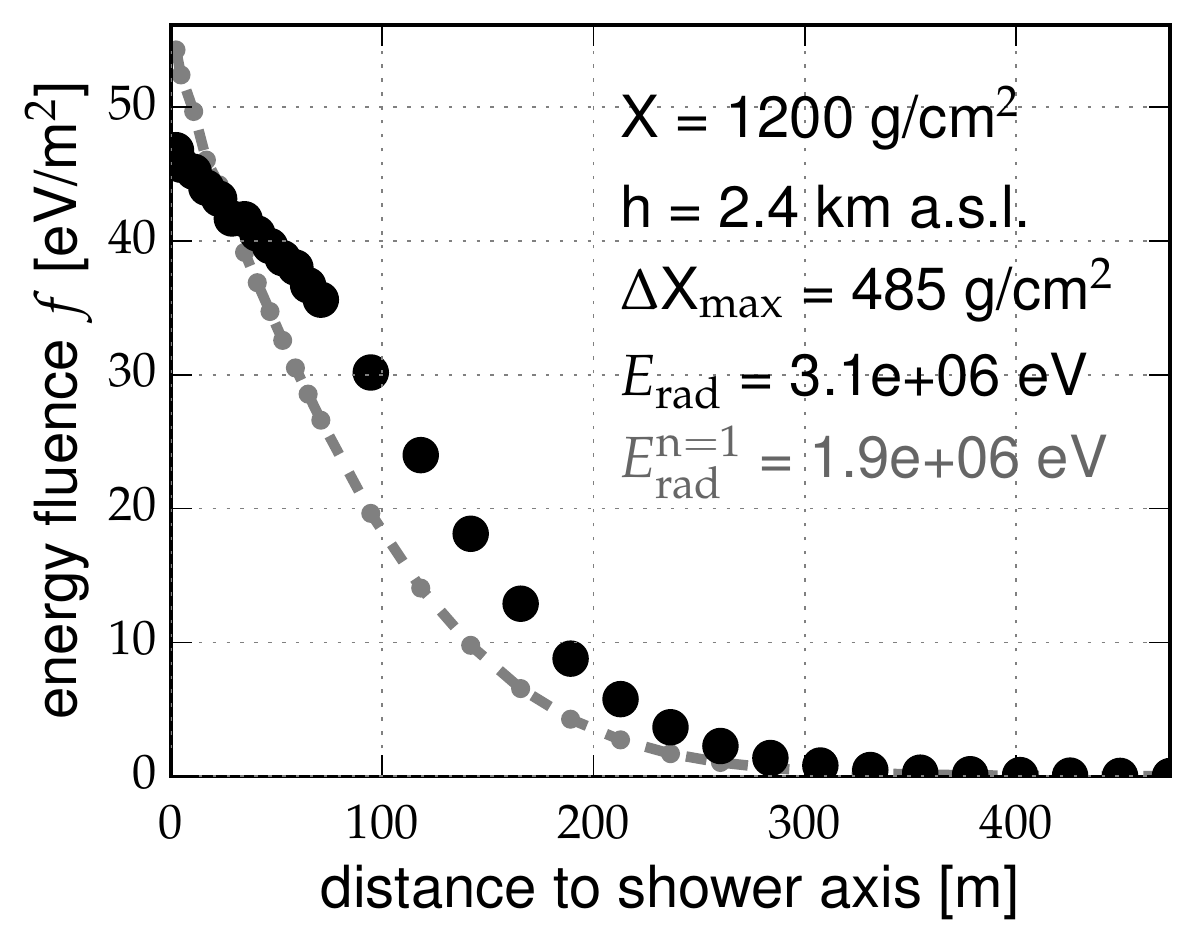}
 \put(\dx, \dy){\Large d)}
 \end{overpic}
 \begin{overpic}[width=0.49\textwidth]{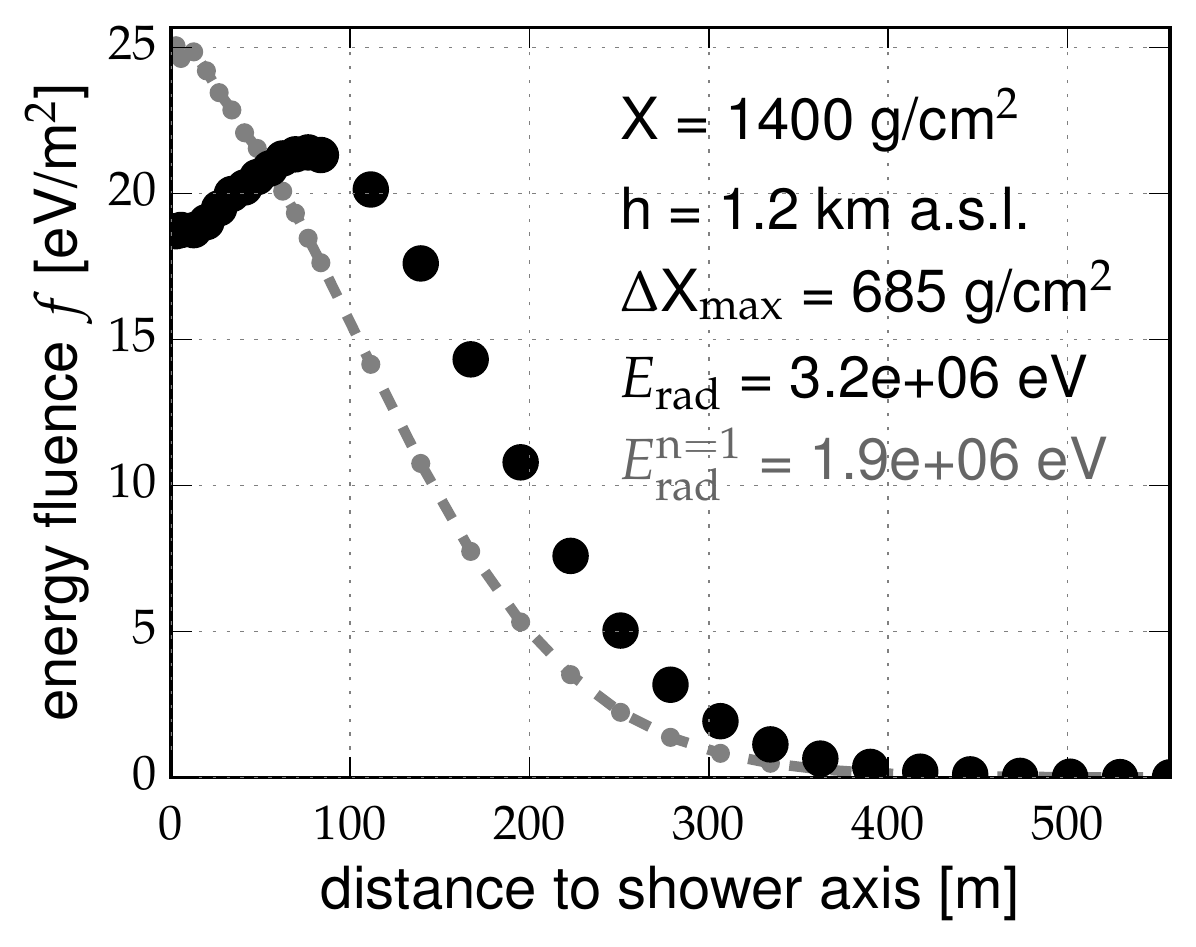}
 \put(\dx, \dy){\Large e)}
 \end{overpic}
 \begin{overpic}[width=0.49\textwidth]{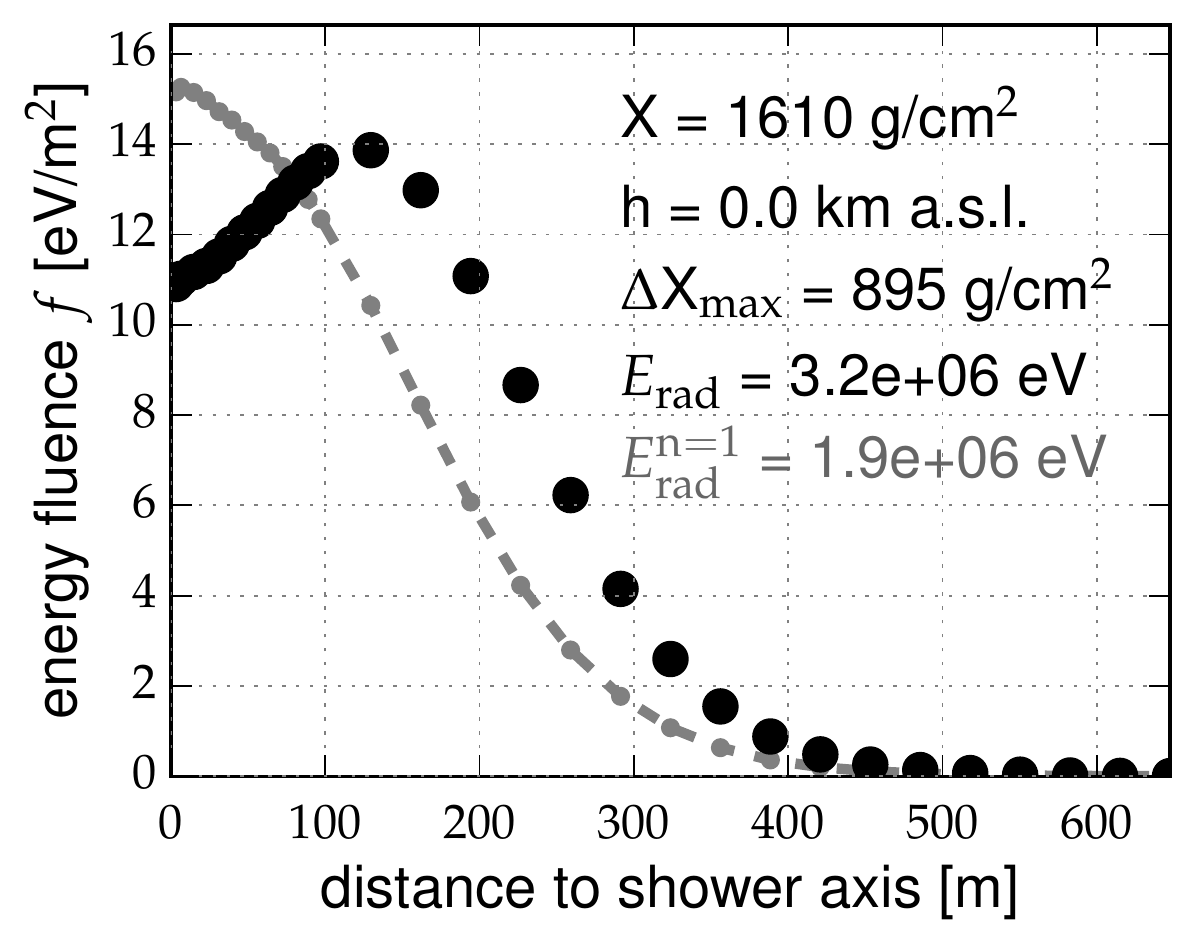}
 \put(\dx, \dy){\Large f)}
 \end{overpic}
 \caption{The black circles show the energy fluence in the \unit[30 - 80]{MHz} band along the positive $\vvB$ axis at different atmospheric depths $X$ for a proton-induced air shower with an energy of \unit[0.5]{EeV} and a zenith angle of 50$^\circ$ which is coming from the south. The dashed gray line shows the energy fluence in case of constant refractivity in the complete atmosphere for the exact same air shower. Please note the different horizontal and vertical axis scales. }
 \label{fig:LDFexample}
\end{figure}

To obtain the radiation energy, we simulate an air shower with observers at different atmospheric depths. Fig.~\ref{fig:LDFexample} shows the lateral distribution of the energy fluence $f$ on the positive $\vvB$ axis for six different atmospheric depths. The second simulated height at $X$ = \unit[700]{g/cm$^2$} (Fig.~\ref{fig:LDFexample}b) is very close the shower maximum $X_\mathrm{max}$ of \unit[715]{g/cm$^2$}. (Throughout the paper we use $X_\mathrm{max}$ to refer to the maximum of the longitudinal profile of the energy deposit $\mathrm{d}E/\mathrm{d}X$ of the electromagnetic part of the air shower.) At the beginning of the shower development, the energy fluence is concentrated around the shower axis and becomes broader for observers at lower altitudes. The largest energy fluence is seen by an observer at the height of the shower maximum $X_\mathrm{max}$. From this height onwards, the total radiation energy continues to increase, but is distributed over a larger area resulting in smaller values of the energy fluence.

From an atmospheric depth of $X$ = \unit[1000]{g/cm$^2$} onwards, the radiation energy basically remains the same whereas the signal distribution becomes increasingly broader. It even occurs that the energy fluence $f$ first increases with an increasing distance from the shower axis before it decreases again. 
This behavior can be explained by Cherenkov-like effects. The refractivity of the atmosphere is not unity, but has a typical value of $n-1 = 2.92 \times 10^{-4}$ at sea level and scales proportionally with air density. For an observer at the Cherenkov angle, all radiation emitted along the shower axis arrives simultaneously at the observer. This leads to additional coherence effects which results in a larger energy fluence at this specific distance from the shower axis. 

To demonstrate the strength of this effect, we resimulated the air shower shown in Fig.~\ref{fig:LDFexample} with a constant refractive index of $n=1$ in the complete atmosphere and show the resulting energy fluence as a dashed line in Fig.~\ref{fig:LDFexample}. The shower development is exactly the same as in the case of realistic modeling of the refractivity. Without Cherenkov-like effects, the energy fluence falls off monotonically with increasing distance to the shower axis and also the resulting radiation energy is significantly lower (cf. Fig.~\ref{fig:LDFexample}).

\begin{figure}[tb]
 \centering
 \includegraphics[width=0.95\textwidth]{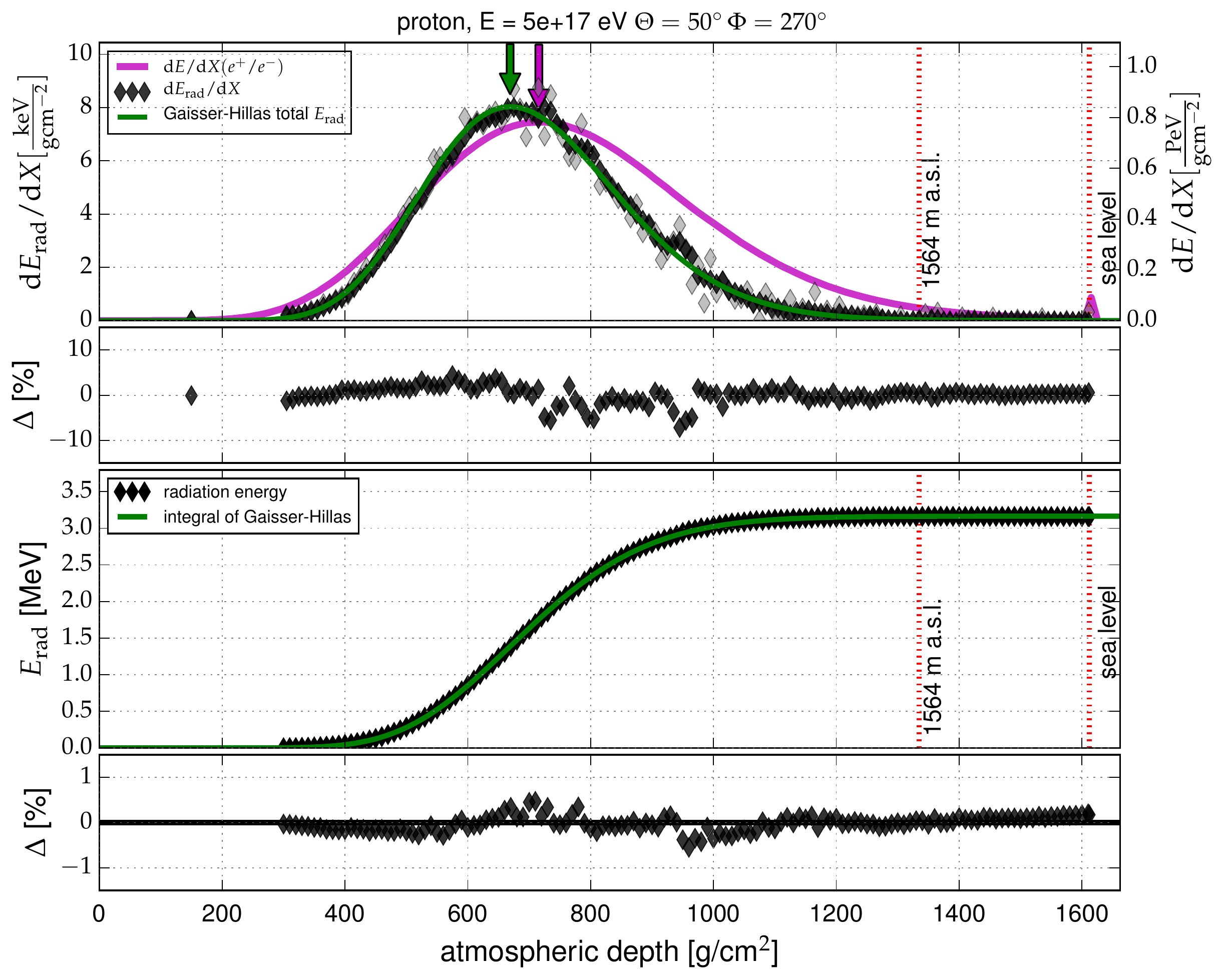}
 \caption{Longitudinal profile of the radiation energy release of the same air shower as shown in Fig.~\ref{fig:LDFexample}. The lower diagram plots the total radiation energy measured at a given atmospheric depth. The green line demonstrates a fit of the integral of a Gaisser-Hillas function to the radiation energy. The residuals are also shown. The upper diagram plots the same data and fit, but as radiation energy release at a given atmospheric depth. Also shown is the energy deposit of the electromagnetic shower particles as a purple line that needs to be read on the right vertical axis. The bump at \unit[$\sim$1600]{g/cm$^2$} comes from shower particles that deposit their energy in the ground. The vertical lines indicate the atmospheric depth that corresponds to the height of the radio detector of the Pierre Auger Observatory and sea level (height of the LOFAR detector) for the simulated geometry. The arrows indicate the position of the shower maximum. See text for details.}
 \label{fig:Long1}
\end{figure}

Using this approach we can investigate the radiation energy and its development with increasing atmospheric depth which is shown in Fig.~\ref{fig:Long1}. The lower part of the figure shows the total radiation energy at a given atmospheric depth, whereas the upper plot shows the radiation energy release per atmospheric depth which is calculated via
\begin{equation}
 \frac{\mathrm{d}E_\mathrm{rad}}{\mathrm{d}X}\left(\frac{X_i + X_{i+i}}{2}\right) = \frac{E_\mathrm{rad}(X_{i+1}) - E_\mathrm{rad}(X_i)}{X_{i+1} - X_i} \, ,
\end{equation}
where $X_i$ is the $i^\mathrm{th}$ simulated height of the observer. In this example the uncertainty of the radiation energy release per atmospheric depth is relatively large, because the difference in radiation energy between two succeeding simulated atmospheric depths is small compared to the uncertainty of $E_\mathrm{rad}$. 
Therefore, we show the running average of $\frac{\mathrm{d}E_\mathrm{rad}}{\mathrm{d}X}$ of five surrounding atmospheric depth bins as black diamonds and the unsmoothed values as gray diamonds. 

In the upper figure we compare the radiation energy release per atmospheric depth, which we will hereafter refer to as the ``longitudinal profile of the radiation energy release'', with the longitudinal profile of the energy deposit of the electromagnetic cascade of the air shower. The profile of the radiation energy release starts with a delay, but reaches its maximum before the maximum of the energy deposit. Thereafter it decreases faster than the energy deposit by the particle shower. 

We note that the radiation energy release $\mathrm{d}E_\mathrm{rad}/\mathrm{d}X$ is conceptually different from the energy deposit of the electromagnetic shower component $\mathrm{d}E/\mathrm{d}X$. The shower particles are absorbed by the atmosphere and deposit their energy in the atmosphere. In contrast, the radiation energy is released in the atmosphere and propagates freely as the atmosphere is transparent for radio emission. 

\subsection{Fit of Gaisser-Hillas Function}
The longitudinal profile of the radiation energy release can be described well with a Gaisser-Hillas function \cite{GaisserHillas} in the version with three parameters. 
\begin{equation}
 \frac{\mathrm{d}E_\mathrm{rad}}{\mathrm{d}X}\left(X\right) = A \left(\frac{X}{X^\mathrm{rad}_\mathrm{max}}\right)^{\frac{X^\mathrm{rad}_\mathrm{max}}{\lambda}} \exp \left(\frac{X^\mathrm{rad}_\mathrm{max}- X}{\lambda}\right) \, ,
 \label{eq:gh}
\end{equation}
where A, $X_\mathrm{max}^\mathrm{rad}$, and $\lambda$ are the free parameters. The integral of Eq.~\eqref{eq:gh} over $X$ gives the total radiation energy, which is the primary objective of this paper. Therefore, we fit the integral of the Gaisser-Hillas function to the total radiation energy measured at a given atmospheric depth (lower panel of Fig.~\ref{fig:Long1}), which we found to result in a more precise determination of the radiation energy. With this approach we can even determine the full radiation energy in cases where the air shower hits the ground before all radiation energy has been emitted. We note that no analytic form of the integral of the Gaisser-Hillas function exists. The integration is therefore performed numerically.

Except for the example shown in Fig.~\ref{fig:Long1}, we simulate 5 - 7 different observation heights per air shower which is sufficient to constrain the Gaisser-Hillas function. We distribute the observation heights around the shower maximum and simulate additional observation heights for air showers that traverse more atmosphere, i.e., air showers that have a large zenith angle\footnote{The first 4 observation heights are distributed symmetrically around the position of the shower maximum {\xmax} at \mbox{$X_\mathrm{max} \pm {\SI{300}{g/cm^2}}$} and \mbox{$X_\mathrm{max} \pm {\SI{50}{g/cm^2}}$} and the 5\textsuperscript{th} observation height is at sea level. If the traversed atmosphere is large enough also an observation level at \mbox{$X_\mathrm{max} + {\SI{500}{g/cm^2}}$} and \mbox{$X_\mathrm{max} + {\SI{1000}{g/cm^2}}$} is added.}.

For all simulated air showers, we compare the parameters of the Gaisser-Hillas function of the radiation energy release with the parameters that describe the energy deposit of the electromagnetic air-shower particles. We find that \xmaxrad is on average \unit[48]{g/cm$^2$} before the \xmax of the energy deposit by the electromagnetic part of the air shower with a scatter of \unit[+6.2/-7.5]{g/cm$^2$}. The difference in the $\lambda$-parameter is \unit[-32]{g/cm$^2$} with a scatter of \unit[4]{g/cm$^2$}. These shifts show no dependence on the shower energy but exhibit a small zenith angle dependence. The difference {\xmaxrad} $-$ {\xmax} shows a constant offset of {\SI{-49}{g/cm^2}} until {60$^\circ$} zenith angle and rises for higher zenith angles to {\SI{-37}{g/cm^2}}. The difference in the {$\lambda$} parameter increases from {\SI{-34}{g/cm^2}} at {0$^\circ$} zenith angle to {\SI{-28}{g/cm^2}} at {80$^\circ$} zenith angle. This zenith dependence is difficult to interpret as it is also influenced by the mixing of geomagnetic and charge-excess radiation that have a different longitudinal profile of the radiation energy release (see Sec.~\ref{sec:longgeoce}).

\subsection{Uncertainties of the total Radiation Energy}
\label{sec:erad_uncertainty}
The dominant uncertainty of the radiation energy originates from the approximation of calculating the radiation energy only from the energy fluence on the positive $\vvB$ axis instead of a full 2D integration of the radio footprint. We estimated the resulting uncertainty by simulating the energy fluence in the entire shower plane for a representative set of shower geometries consisting of 150 air showers. We compared the radiation energy calculated from the energy fluence on the positive $\vvB$ axis with the radiation energy determined via a full two-dimensional integration of the energy fluence. We find a scatter of 3\% with a bias of 3\% towards higher radiation energies, i.e., in our approach we overestimate the radiation energy by 3\% on average.

The reason for the observed asymmetry that results in the bias is well understood and can be attributed to two effects. As discussed before, we overestimate the radiation energy by $\sim$1\% by the approximation that the geomagnetic and charge-excess components of the electric field are in-phase. In addition, the Cherenkov-like effects that lead to additional coherence and thereby larger energy fluences are not radially symmetric around the shower axis. Except for exactly vertical showers, the observer positions in the shower plane have different heights above ground. 
Therefore, observer positions with the same distance to the shower axis see different effective refractivities on their line of sight to the emission region.
The average refractive index between a position below the shower axis and the emission region is larger than for a position above the shower axis with the same distance to the shower axis. And as already visible from Fig.~\ref{fig:LDFexample}, larger refractivities generally result in more radiation energy. Hence, depending on the orientation of the positive $\vvB$ axis in ground coordinates, we over or underestimate the radiation energy. The orientation of the positive $\vvB$ axis depends on the incoming direction of the air shower and therefore the bias in the determination of the radiation energy exhibits directional dependence. For the distribution of incoming directions in our simulated data set, the positive $\vvB$ axis is mostly below the shower axis, resulting in an additional overestimation of the radiation energy of 2\% on average. The scatter of 3\% results from the dependence on the shower geometry. 
However, this result is not critical for the desired accuracy of this analysis. In the following, we will correct the radiation energy by 3\% and use the 3\% scatter as uncertainty of the radiation energy. 

An additional source of uncertainty arises from the determination of the radiation energy on the positive $\vvB$ axis itself. We use the tiny scatter of the radiation energy around the integral of the Gaisser-Hillas function (cf. Fig.~\ref{fig:Long1} bottom) to estimate this uncertainty. The scatter can have two reasons: either the Gaisser-Hillas function in the version with three parameters does not describe the longitudinal profile of the radiation energy release in all detail, or the scatter is due to statistical uncertainties which originate from the determination of the simulated radiation energy and/or from fluctuations in the simulation itself. We use this scatter to estimate the precision of the radiation energy determination. Inserting an uncertainty of 0.5\%, the distribution of the $\chi^2$ probability of the fit of the integral of the Gaisser-Hillas function of all simulated air showers is distributed uniformly. For a slightly larger uncertainty (0.7\%) the $\chi^2$ probability distribution indicates an overestimation of the errors. Furthermore, we observe that the Gaisser-Hillas function fits slightly better for air showers with proton primaries than for air showers with iron primaries.

Using this upper limit of 0.5\% as an estimate for the statistical uncertainty of the radiation energy that has been released up to a specific observation height, we determine the uncertainty of the total radiation energy, i.e., the integral over the fitted Gaisser-Hillas function, in a dedicated toy Monte Carlo. If the radiation energy is released fully in the atmosphere, the uncertainty is 0.3\%. As long as more than 90\% of the radiation energy is released in the atmosphere, the uncertainty is less than 1\%, and even if only 70\% of the radiation energy is released in the atmosphere, the uncertainty is still below 2\% without any bias.

\newpage

\section{Decomposition of the Radiation Processes}
\label{sec:longgeoce}

\begin{figure}[ptb]
 \centering
 \vspace{30px}
 \includegraphics[width=1\textwidth]{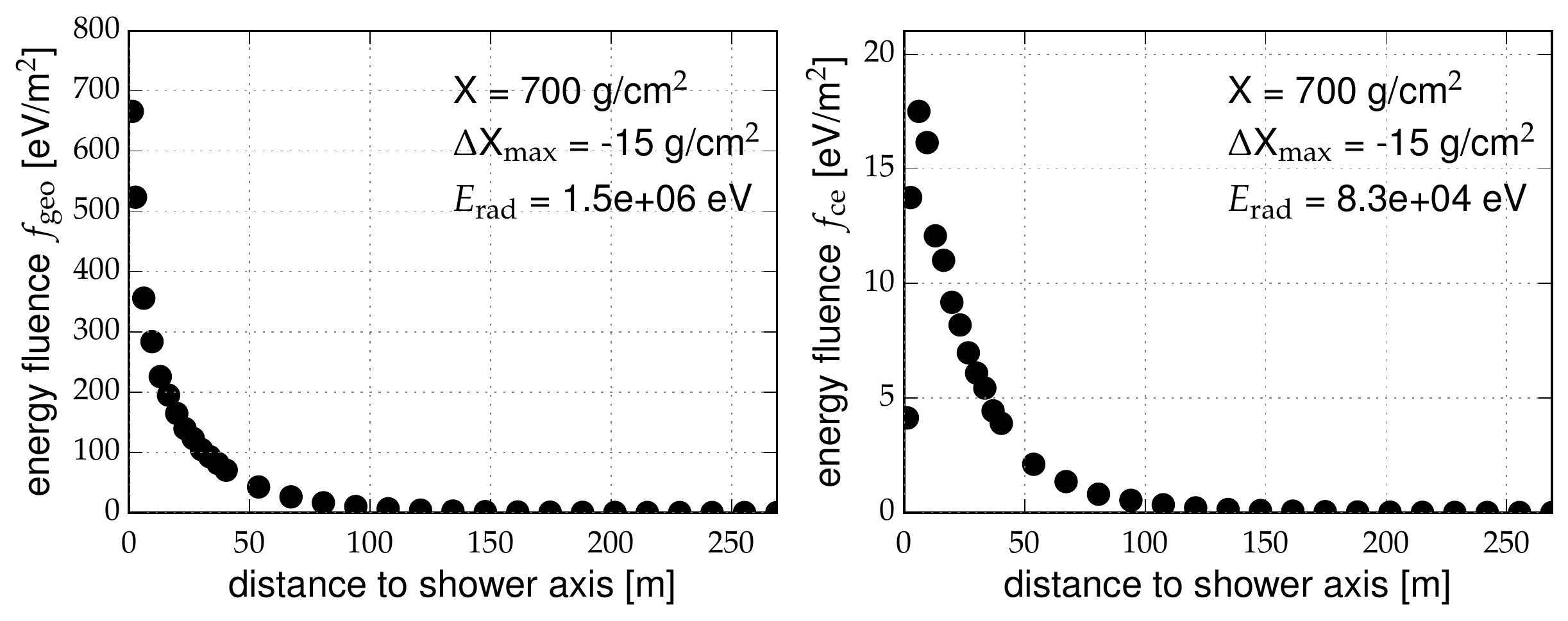}
 \includegraphics[width=1\textwidth]{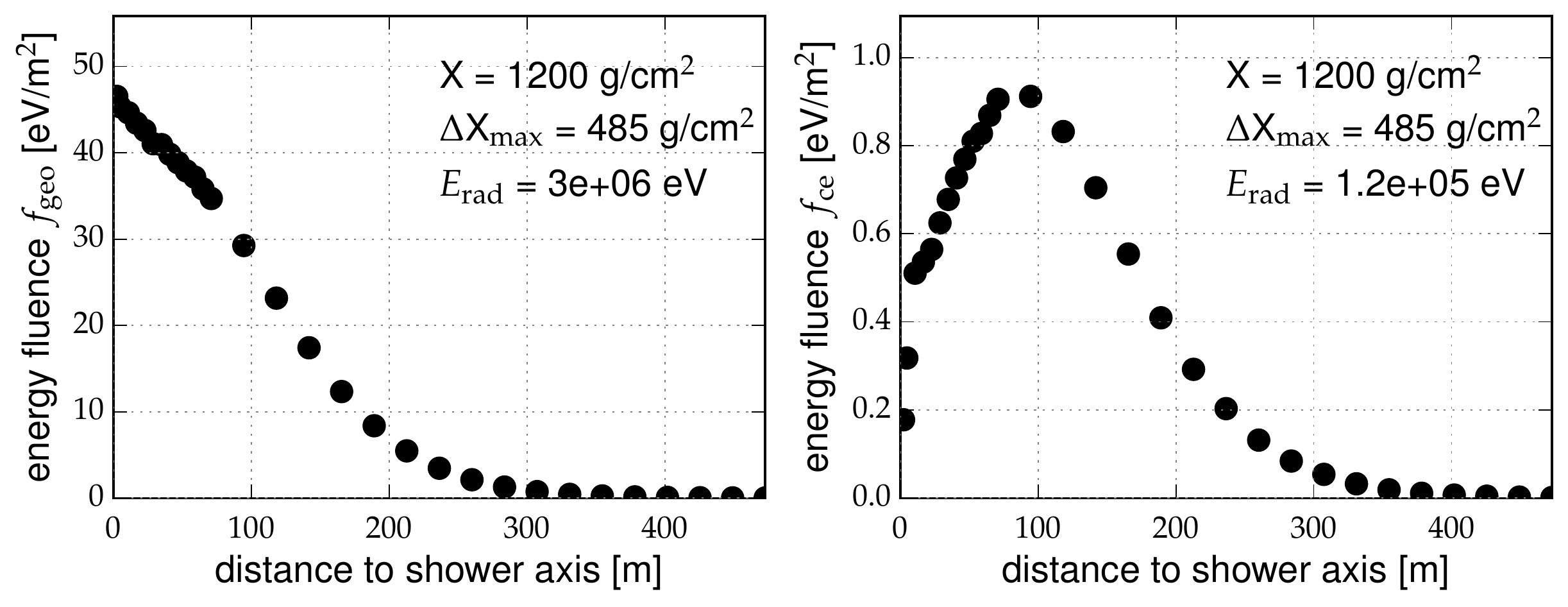}
 \includegraphics[width=1\textwidth]{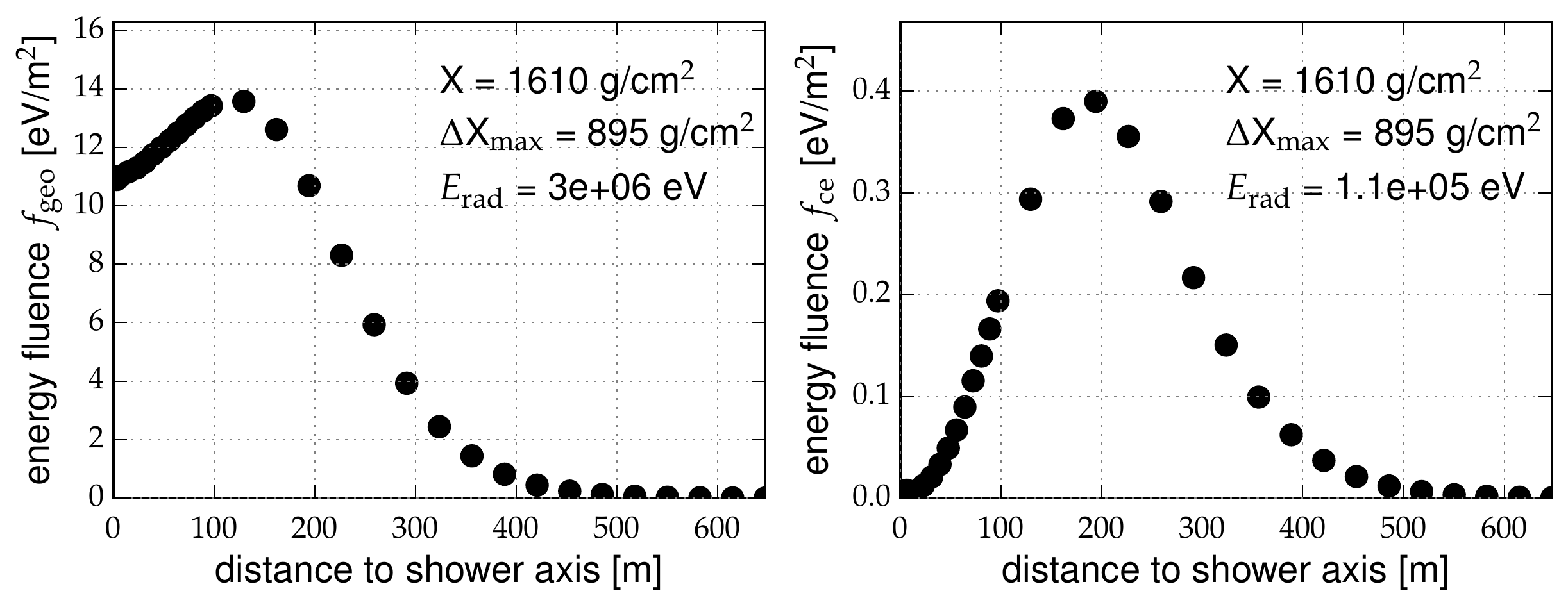}
 \caption{Distribution of the energy fluence in the \unit[30 - 80]{MHz} band along the positive $\vvB$ axis at different atmospheric depths for the same air shower as shown in Fig.~\ref{fig:LDFexample}. Here, the total energy fluence is split into the part that originates from the geomagnetic emission (plots on the left side) and the part that originates from the charge-excess emission (plots on the right side). Please note the different vertical axis scales.}
 \label{fig:LDFexample_geoce}
\end{figure}

Our method to extract the radiation energy from the air shower simulations allows a decomposition into the part originating from the geomagnetic emission \eradgeo and the part originating from the charge-excess emission \eradce. Fig.~\ref{fig:LDFexample_geoce} shows the lateral distribution of the energy fluence of the geomagnetic and the charge-excess component at three different atmospheric depths. The shape of the lateral signal distribution of the two components is different. The geomagnetic LDF has the same shape as the LDF of the total energy fluence (cf. Fig.~\ref{fig:LDFexample}), because the geomagnetic emission is the dominant mechanism in this example. 

The energy fluence of the charge-excess component is zero at the shower axis, then increases with increasing distance before it decreases again.
The further away the observer is from the shower maximum, the more prominent this effect is. This behavior is expected, as the charge-excess emission is polarized radially towards the shower axis. This implies that the polarization of the charge-excess emission turns by 180$^\circ$ when going from one side of the shower to the other, which can only be a continuous transition if the amplitude crosses zero as well. 

\begin{figure}
 \centering
 \includegraphics[width=0.95\textwidth]{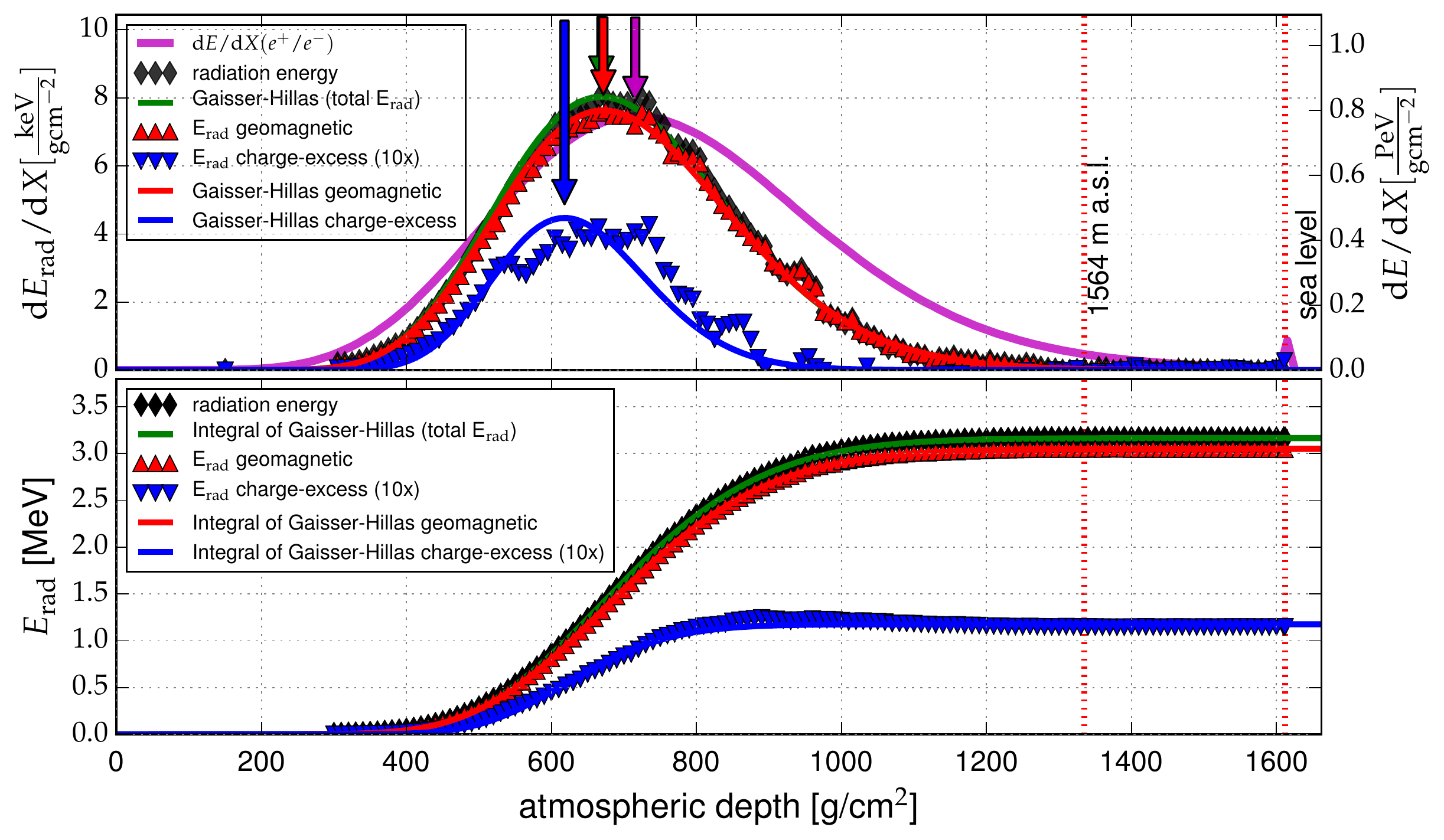}
 \caption{Longitudinal profile of the radiation energy release for the same data as shown in Fig.~\ref{fig:Long1}. Here also the parts of the radiation energy release with geomagnetic and charge-excess origin are shown. The charge-excess radiation energy is multiplied by a factor of 10.}
 \label{fig:Long2}
\end{figure}

From the geomagnetic and charge-excess LDFs we calculate the radiation energies at different atmospheric depths to obtain the longitudinal profiles of the radiation energy release of the two emission mechanisms (see Fig.~\ref{fig:Long2}). 
The longitudinal profile of the geomagnetic radiation energy release follows the profile of the total radiation energy release as the amount of charge-excess radiation energy is small for this air shower.

The total charge-excess radiation energy exhibits some unexpected features. After the radiation energy has reached its maximum at $\sim$\unit[900]{g/cm$^2$}, it decreases slightly with increasing atmospheric depths and then remains constant. 
A plausible explanation is given by the fact that the orientation of the radial electric-field vectors associated with charge-excess radiation reverses when the total net charge excess reaches its maximum and thus turns from a growing net charge to a diminishing net charge. The radiation of the late stages of the air shower evolution thus interferes destructively
with that of the early stages, leading to a decrease of the total, integrated radiation energy. The decrease in radiation energy means that energy has to be absorbed by some mechanism. This could, for example, happen by absorption of the electromagnetic radiation by charged particles in the shower cascade. As the total amount of energy that needs to be absorbed is more than ten orders of magnitude smaller than the shower energy it is irrelevant for the shower development and is therefore not implemented in the simulation. Another possible explanation is that the polarization changes slightly during the shower development, so that a part of the geomagnetic radiation is also visible in the $\vvB$ polarization or vice versa. In addition, the longitudinal profile of the charge-excess radiation energy release deviates from a Gaisser-Hillas function. Therefore, the results on \xmaxrad obtained from the Gaisser-Hillas fit must be interpreted with caution; however, as the integral of the Gaisser-Hillas function is fitted to the cumulative energy distribution, the radiation energy is still estimated correctly.

For all simulated air showers, we fit the integral of a Gaisser-Hillas function separately to the geomagnetic and charge-excess radiation energy at a given atmospheric depth and compare the corresponding Gaisser-Hillas function (cf. Eq.~\eqref{eq:gh}) to the longitudinal profile of the electromagnetic part of the air shower. For the geomagnetic radiation energy we find $X_\mathrm{max}^\mathrm{rad} - X_\mathrm{max} = \unit[-47^{+6}_{-8}]{g/cm^2}$ and $\Delta \lambda = \unit[(-32 \pm 4)]{g/cm^2}$. 
For the charge-excess radiation energy we find a significantly different behavior, with $X_\mathrm{max}^\mathrm{rad} - X_\mathrm{max} = \unit[-87^{+13}_{-10}]{g/cm^2}$ and $\Delta \lambda = \unit[(-48^{+6}_{-7})]{g/cm^2}$ under the assumption that the longitudinal profile of the radiation energy release with charge-excess origin can be described with a Gaisser Hillas function.

With the separate determination of \eradgeo and \eradce we are able to generalize previous work regarding the relative charge-excess strength $a$ \cite{VriesScholtenWerner2013, AERAPolarization, LofarPolarization2014, Kostunin2015}, where $a$ was defined as the ratio of the electric field amplitudes of the charge-excess and geomagnetic components for maximum geomagnetic emission (\sina = 1). In \cite{AERAPolarization}, an average value of $a$ = 14\% was measured at the Auger site. Using LOFAR \cite{LofarPolarization2014}, additional dependencies of $a$ on the distance to the shower axis and zenith angle were found. The dependence of $a$ on the distance from the shower axis can be explained by the different shapes of the geomagnetic and charge-excess LDFs (cf. Fig.~\ref{fig:LDFexample_geoce}). 

In the following we generalize the definition of the charge-excess fraction using the radiation energy, i.e., the integral of the LDF, and define
\begin{equation}
 a = \sin \upalpha \, \sqrt{\frac{E_\mathrm{rad}^\mathrm{ce}}{E_\mathrm{rad}^\mathrm{geo}}} \, ,
\end{equation}
where \eradgeo and \eradce are the radiation energies after the shower has fully developed, which corresponds to the integrals of the respective Gaisser-Hillas functions. We do not study the development of the charge-excess fraction during the shower development.  We take the square root of the ratio of radiation energies to be consistent with the previous work \cite{AERAPolarization, LofarPolarization2014}, because the radiation energy scales quadratically with the electric field amplitude. 

\begin{figure}[tb]
 \centering
 \includegraphics[width=0.85\textwidth]{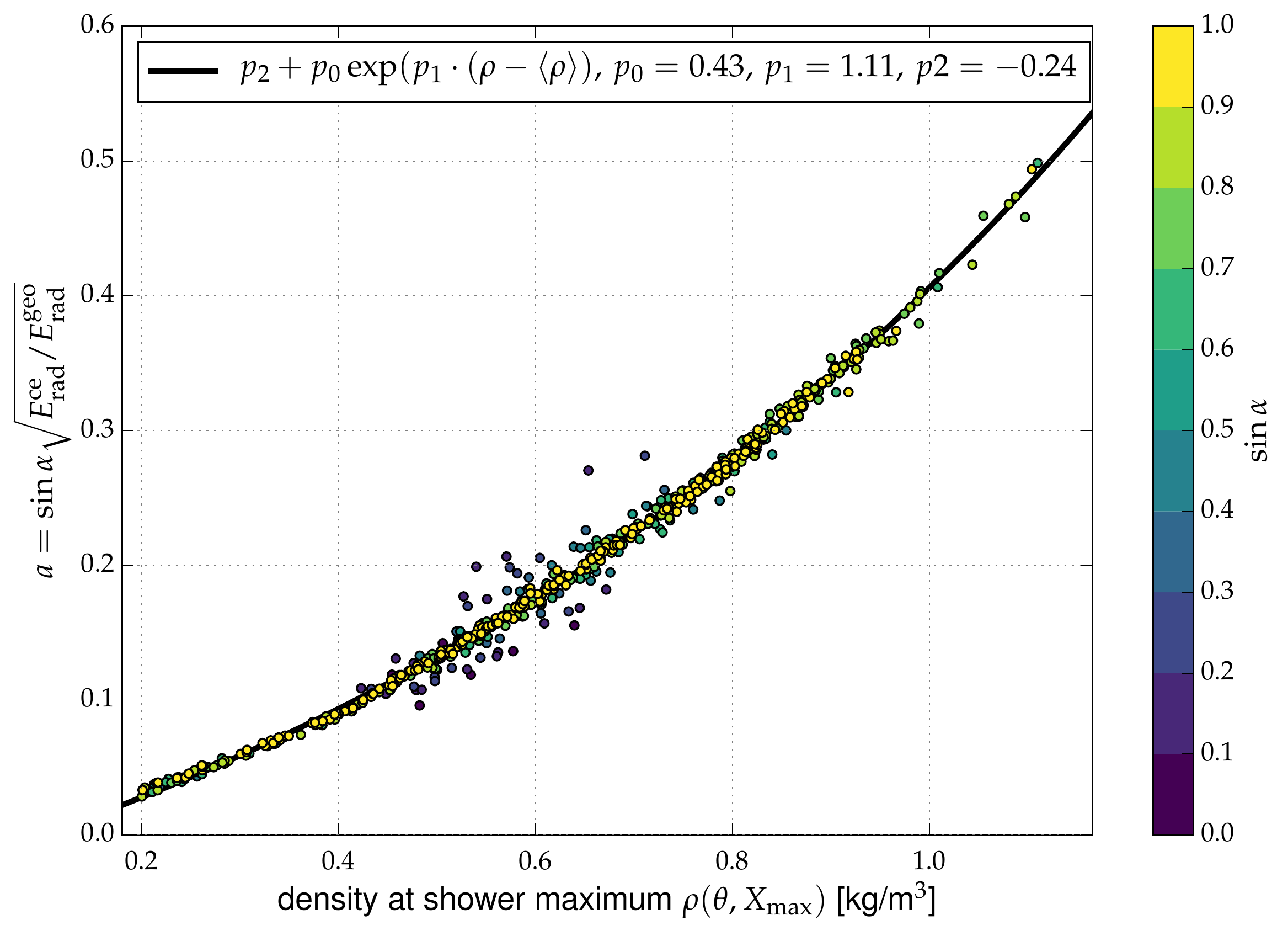}
 \caption{Dependence of charge-excess fraction $a$ on the atmospheric density at the position of the shower maximum. The dashed line represents an exponential fit to the data.}
 \label{fig:adxmax}
\end{figure}

We find that the value of $a$ depends on where in the atmosphere the radiation is generated. More precisely, $a$ depends on the density in the part of the atmosphere in which the radiation is generated. As most of the radiation energy is released near the shower maximum \xmax, we use the atmospheric density \rhoxmax at the position of the shower maximum to parametrize this dependence. The density \rhoxmax depends on the zenith angle $\theta$ and \xmax of the air shower and can be calculated from the atmospheric density profile. We use the model of the US standard atmosphere as it is also used in the air-shower simulations.

Fig.~\ref{fig:adxmax} shows the dependence of $a$ on the atmospheric density at the shower maximum. We observe that $a$ increases with increasing \rhoxmax which is consistent with the result of \cite{LofarPolarization2014} who reported an increase of $a$ with decreasing zenith angle. We do not observe any difference between air showers induced by protons or iron nuclei. We describe the dependence with an exponential function with the following parameters:

\begin{equation}
 a(\rho_\mathrm{X_\mathrm{max}}) = 0.43  \left(e^{\unit[1.11]{m^3/kg} \, (\rho_\mathrm{X_\mathrm{max}} - \langle \rho \rangle)}\right) - 0.24\, ,
 \label{eq:adxmax}
\end{equation}
where $\langle \rho \rangle$ = \unit[0.65]{kg/m$^3$} is the atmospheric density at the shower maximum for an average zenith angle of 45$^\circ$ and an average $\langle X_\mathrm{max} \rangle$ = \unit[669]{g/cm$^2$} as predicted for a shower energy of \unit[1]{EeV} for a 50\% proton/50\% iron composition using QGSJetII-04 \cite{Domenico2013}.
The color code of Fig.~\ref{fig:adxmax} shows that only air showers with small values of \sina exhibit some deviation from the parametrization.

Using Eq.~\eqref{eq:adxmax} together with the geometry of the air shower and a description of the atmosphere, the ratio of $E_\mathrm{rad}^\mathrm{ce}/E_\mathrm{rad}^\mathrm{geo}$ can be converted to the position of the shower maximum. Therefore in a measurement, a separate determination of the geomagnetic and charge-excess radiation energies, e.g., through a combined fit of the geomagnetic and charge-excess LDFs according to Eq.~\eqref{eq:uvvB} and \eqref{eq:uvB}, can be used to estimate the position of the shower maximum \xmax which is an estimator of the mass of the cosmic ray.

\newpage

\section{Properties of the Radiation Energy}

\subsection{Dependence on Shower Energy}
The radiation energy originates from the radiation generated by the electromagnetic part of the air shower. Hence, the radiation energy correlates best with the energy of the electromagnetic cascade and not with the complete shower energy, which includes energy carried by neutrinos and high-energy muons that are not relevant for radio emission. This is not necessarily a disadvantage but a benefit for several analyses. Other detection techniques are also only sensitive to the electromagnetic shower energy, such as fluorescence or air-Cherenkov telescopes, which allows for a direct cross calibration. Also the measurement of the electromagnetic shower energy can be combined with a separate measurement of the muonic shower component (e.g. via buried scintillators) to obtain a sensitivity on the particle species of the primary cosmic-ray particle. If, however, the goal is to measure the cosmic-ray energy, an additional scatter arises from the varying fraction of cosmic-ray energy entering the electromagnetic cascades for different particle species. For a cosmic-ray energy of {\SI{1}{EeV}} the difference in electromagnetic shower energy is $\sim$4.5\% for the two extreme cases of cosmic-rays being protons or iron nuclei and decreases to 3\% at {\SI{10}{EeV}} \cite{Engel:2011zzb}. The average fraction of the electromagnetic shower energy can be inferred from hadronic interaction models assuming a specific cosmic-ray composition or using a direct measurement as presented in {\cite{InvisibleEnergy2013}}.

Before correlating the radiation energy with the energy of the electromagnetic cascade, the geomagnetic radiation energy needs to be corrected for geometry. The radiation energy of the geomagnetic contribution depends on the magnitude of the geomagnetic field $B_\mathrm{Earth}$ and the angle $\upalpha$ between the shower direction and the geomagnetic field and scales with $\sin^2 \upalpha$. We first study only the dependence on $\sin^2 \upalpha$ as all simulations were performed using the same geomagnetic field strength. We investigate the scaling with the geomagnetic field in the next section. In contrast, the radiation energy of the charge-excess component does not depend on the geomagnetic field. 
From Eq.~\eqref{eq:geocedec} it follows that the radiation energy is the sum of the radiation energy of the geomagnetic emission and the radiation energy of the charge-excess emission \cite{Kostunin2015}. We therefore correct the radiation energy by
\begin{equation}
 S_\mathrm{RD} = \frac{E_\mathrm{rad}}{a(\rho_\mathrm{X_\mathrm{max}})^2 + (1-a(\rho_\mathrm{X_\mathrm{max}})^2) \sin^2\upalpha } \, ,
 \label{eq:sradio1}
\end{equation}
where $a(\rho_\mathrm{X_\mathrm{max}})$ is the parametrization of the charge-excess fraction from Eq.~\eqref{eq:adxmax}.
We fit a power law of the form
\begin{equation}
\label{eq:powerlaw}
 S_\mathrm{RD} = A \times \unit[10^7]{eV} \,(E_\mathrm{em}/\unit[10^{18}]{eV})^B
\end{equation}
and find that $S_\mathrm{RD}$ scales quadratically with the energy of the electromagnetic cascade $E_\mathrm{em}$ as expected for coherent emission. The scatter around the calibration curve is less than 10\%. 

In addition to this $\sin \upalpha$ correction we expect a second-order dependence.
The air shower develops according to the slant depth, whereas the amount of radiation increases with the geometric path length of the shower development. In a region of low atmospheric density, the ratio of the geometric path length and propagation length measured in atmospheric depth is larger than in high-density regions. 
Therefore, a shower that develops early in the atmosphere has a slightly larger radiation energy than a shower with the same energy that develops deeper in the atmosphere, as the air density increases with increasing atmospheric depth. This depends primarily on the zenith angle of the incoming direction of the air shower and secondarily on the position of the shower maximum. We again use the atmospheric density at the position of the shower maximum \rhoxmax to parametrize this second-order dependence.

\begin{figure}[tb]
 \centering
 \includegraphics[width=0.85\textwidth]{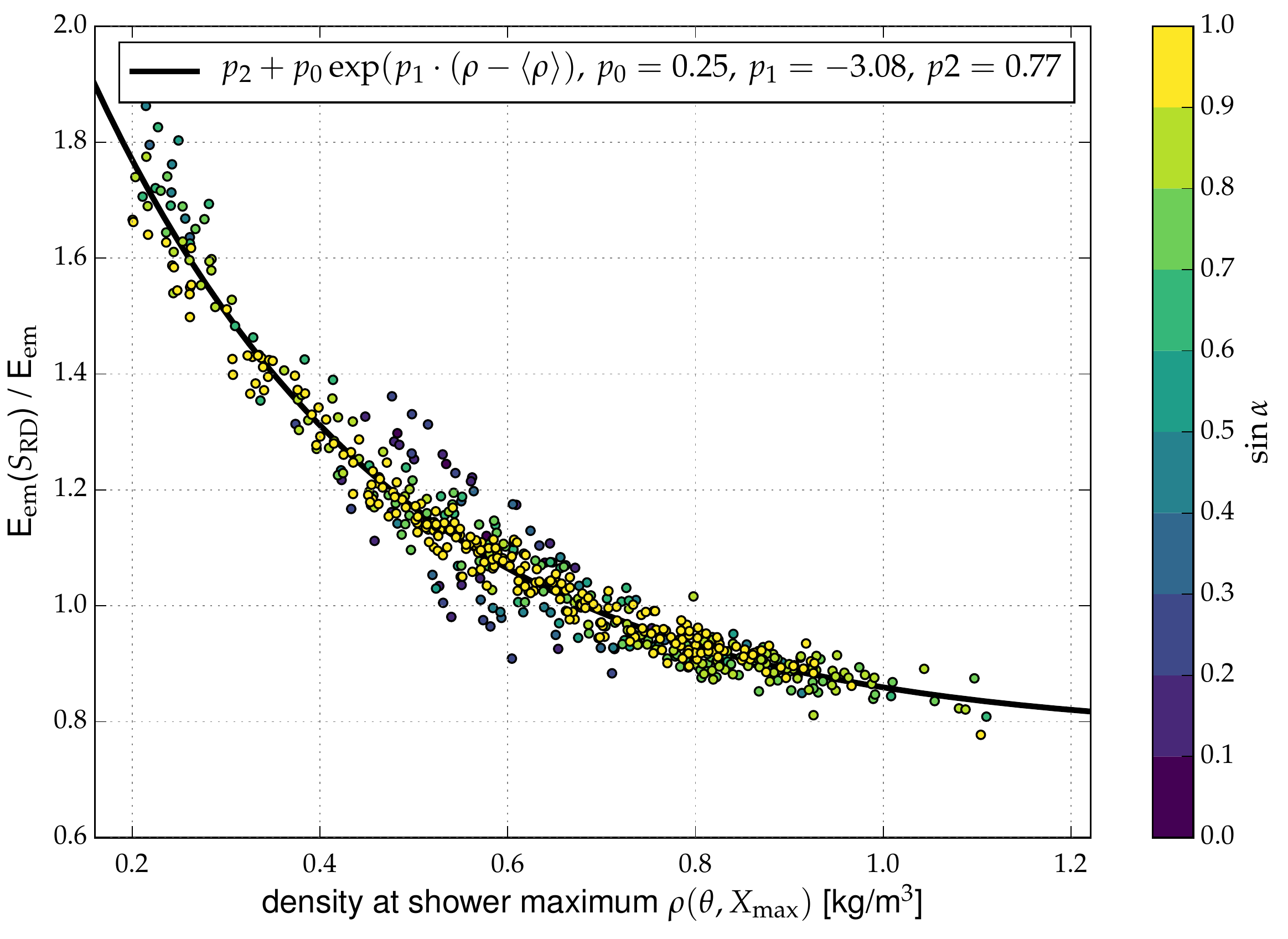}
 \caption{Residuals of the correlation between $S_\mathrm{RD}$ and the energy in the electromagnetic part of the air shower plotted as a function of the density at the shower maximum. An exponential function is adapted to the data.}
 \label{fig:energyres_dxmax}
\end{figure}

The residuals of the relation between the corrected radiation energy $S_\mathrm{RD}$ as defined in Eq.~\eqref{eq:sradio1} and the energy in the electromagnetic part of the air shower indeed show a dependence on \rhoxmax and are shown in Fig.~\ref{fig:energyres_dxmax}. For air showers with the same electromagnetic shower energy, the radiation energy increases with decreasing atmospheric density \rhoxmax. We incorporate this dependence of the radiation energy by applying a second correction term to $S_\mathrm{RD}$:

\begin{equation}
S_\mathrm{RD}^{\rho} = \frac{E_\mathrm{rad}}{a(\rho_\mathrm{X_\mathrm{max}})^2 + (1-a(\rho_\mathrm{X_\mathrm{max}})^2) \sin^2\upalpha} \,\, \frac{1}{\left(
1 - p_0 + p_0 \,  \exp[p_1 (\rho_{X_\mathrm{max}} - \langle \rho \rangle)]\right)^2 } \,.
 \label{eq:Srddxmax}
\end{equation}
As in Eq.~\eqref{eq:adxmax} we normalize the correction to an average $\langle \rho \rangle$ = \unit[0.65]{kg/m$^3$}.

\begin{figure}[tb]
 \centering
 \includegraphics[width=0.85\textwidth]{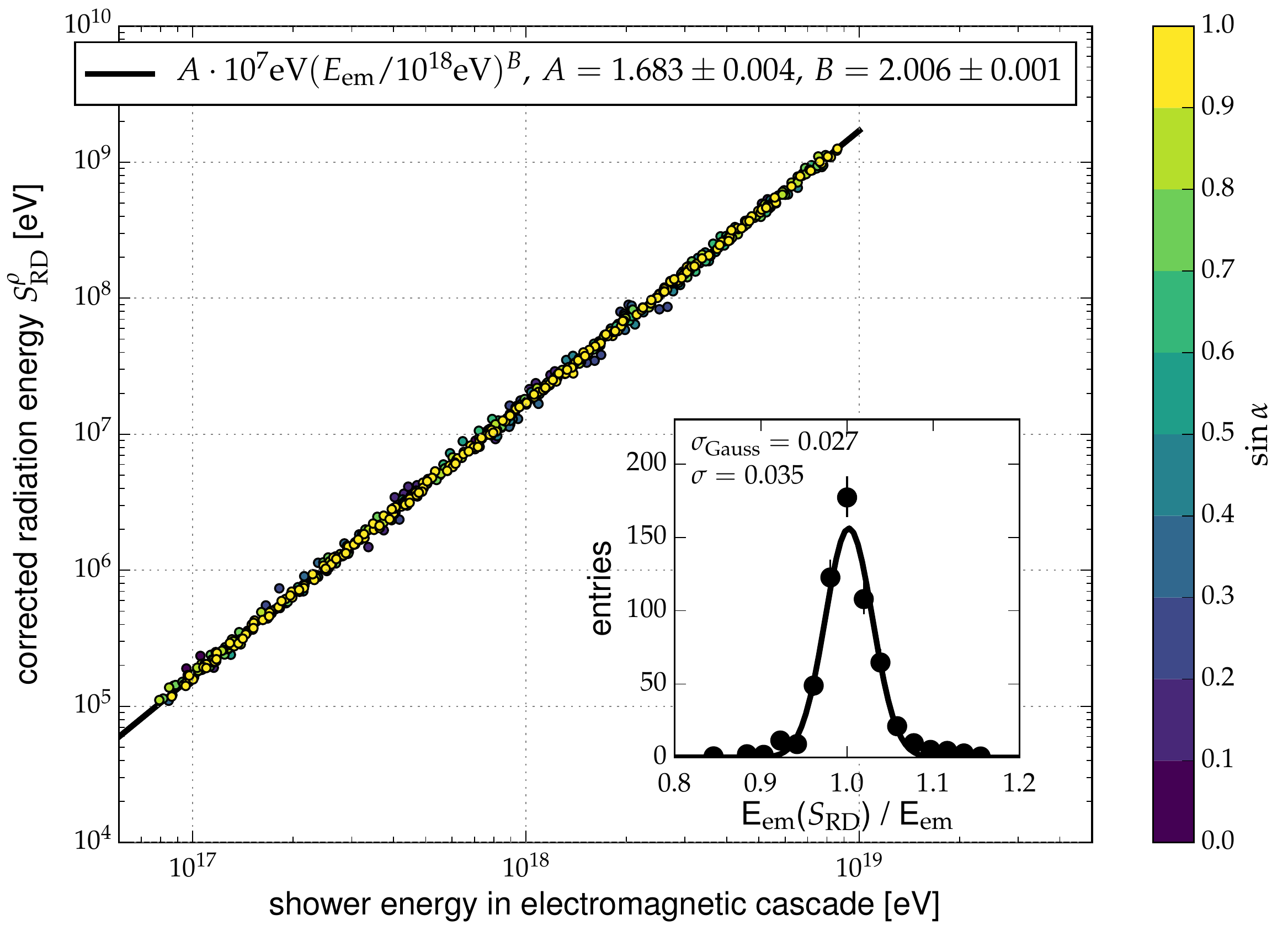}
 \caption{Correlation between the energy in the electromagnetic part of the air shower and the radiation energy corrected for $\sin \upalpha$ as well as the atmospheric density at the shower maximum (cf. Eq.~\eqref{eq:Srddxmax}). The black line shows a power-law fit to the data. The inset figure shows the scatter around the calibration curve. The color scale shows the value of $\sin \upalpha$. }
 \label{fig:energycalib_dxmax}
\end{figure}

The correlation between the corrected radiation energy $S_\mathrm{RD}^{\rho}$ and the energy in the electromagnetic part of the air shower is shown in Fig.~\ref{fig:energycalib_dxmax}. 
We determine the calibration constants $A$ and $B$ of Eq.~\eqref{eq:powerlaw} and the parameter $p_0$ and $p_1$ in a combined chi-square fit and present the results in Tab.~\ref{tab:srddxmax}. 

\begin{table}[h]
\centering
\begin{tabular}{cr@{ $\pm$ }l}
 \hline \hline
 $A$ & 1.683 & 0.004 \\
 $B$ & 2.006 & 0.001 \\
 $p_0$ & 0.251 & 0.006 \\
 $p_1$ & -2.95 & 0.06 \unit{m$^3$/kg} \\
 \hline
 \hline
\end{tabular}
\caption{Best-fit parameters for an energy estimator with density at shower maximum correction as defined in Eq.~\eqref{eq:Srddxmax}.}
\label{tab:srddxmax}
\end{table}

The reduced chi-squared is $1.4$, which may indicate an even higher-order dependence of the radiation energy not taken into account in Eq.~\eqref{eq:Srddxmax}. Hence, the uncertainties of the fit parameters are also slightly underestimated. 
The parameters $p_0$ and $p_1$ exhibit a strong correlation which is, however, not critical as the parameters are always used mutually in the \rhoxmax correction and are never inspected separately. In addition, parameter $A$ also shows some correlation with $p_0$ and $p_1$ as $A$ depends on the absolute normalization of the \rhoxmax correction. To estimate a possible bias we fixed the parameters $p_0$ and $p_1$ to the values obtained in the separate fit of Fig.~\ref{fig:energyres_dxmax} and find a value of $A$ that is well compatible with the result of the combined chi-square fit within the uncertainties.

The inset of Fig.~\ref{fig:energycalib_dxmax} shows a histogram of the scatter around the calibration curve. The standard deviation as computed from the data amounts to 3.5\% owing to tails of the distribution which are related to data points with small values of \sina that exhibit larger fluctuations. If we fit a Gaussian distribution to the binned data we obtain an uncertainty of 2.7\%. The color code of the data points in Fig.~\ref{fig:energycalib_dxmax} shows that the fluctuations are not biased to a certain direction, e.g., towards smaller radiation energies, but are instead centered around the calibration curve. Hence, we conclude that our method works for all incoming directions with slightly larger uncertainties for showers with small values of $\sin \upalpha$.

Furthermore, we investigated possible differences between air showers with proton and iron primaries after we applied the corrections.
The difference in the fraction of cosmic-ray energy that enters the electromagnetic cascade is irrelevant as we correlate $S_\mathrm{RD}^{\rho}$ directly with the electromagnetic energy. As we also correct for the density at the shower maximum, our analysis is insensitive to the difference in the average \xmax between different particle species. However, if we analyze only air showers induced by protons (induced by iron nuclei), we find a 2\% larger (2\% smaller) calibration constant $A$, i.e., proton-induced air showers have a larger corrected radiation energy \srdxmax than iron-induced air showers. As the radiation energy scales quadratically with the electromagnetic shower energy, the $\pm$2\% shift in radiation energy corresponds to a $\pm$1\% shift in the electromagnetic shower energy, which is well below current experimental uncertainties. Hence, the slight dependence of the corrected radiation energy on the cosmic-ray mass is irrelevant for practical purposes.

\subsection{Scaling with the Geomagnetic Field}
\label{sec:bfield}
The amount of radiation energy that is released by an air shower depends on the geomagnetic field. More precisely, the contribution of the geomagnetic emission process to the radiation energy scales with the magnitude of the geomagnetic field $B_\mathrm{Earth}$. Naively, one would assume that the amplitude of the electric-field pulse scales proportionally with the magnitude of the Lorentz force which is proportional to the magnitude of the geomagnetic field. The radiation energy of the geomagnetic contribution would thus be proportional to the magnitude of the geomagnetic field squared. However, we investigated the scaling with the geomagnetic field in the air-shower simulations and found that the radiation energy of the geomagnetic component deviates significantly from a quadratic scaling with $B_\mathrm{Earth}$ and scales approximately with $(B_\mathrm{Earth})^{1.8}$.

We obtained this number by simulating air showers with different magnitudes of the geomagnetic field. For each air shower we calculated the corrected radiation energy according to Eq.~\eqref{eq:Srddxmax} and corrected the geomagnetic part not only by $\sin^2 \alpha$, but also by $(B_\mathrm{Earth}/\unit[0.243]{G})^k$ and determined the scaling factor $k$ that best fits the simulations. The value of \unit[0.243]{G} is the geomagnetic field strength at the location of the Pierre Auger Observatory and is used as a reference in this analysis. We simulated air showers with geomagnetic field magnitudes ranging from the reference field strength to three times the reference field strength which covers the complete range of variations in the geomagnetic field strength occurring on Earth.

A plausible explanation for the deviation from the naively expected quadratic scaling is that the electrons and positrons do not propagate unimpeded but interact continuously with air molecules. The geomagnetic emission should therefore not be confused with synchrotron emission. The shower particles move chaotically within the shower front, they collide with other particles, get accelerated and deaccelerated and they drift only on average into the direction of the Lorentz force. The drift velocity is thereby much smaller than the average velocity of the particles inside the shower front \cite{Scholten200894}. With stronger acceleration by a larger geomagnetic field also the number of collisions with other particles increases resulting in a slightly smaller effective acceleration.

Within the variations of the geomagnetic field strength occurring on Earth and the desired accuracy of this analysis, the scaling with the geomagnetic field can be described effectively with $(B_\mathrm{Earth})^{1.8}$. However, we note that we observe a small second order dependence. 
The gain in radiation energy, if the geomagnetic field is increased from $B = B_\mathrm{Auger}$ to $B = 1.5 \times B_\mathrm{Auger}$, is slightly larger than the gain in radiation energy, if the geomagnetic field is increased from $B = 2 \times B_\mathrm{Auger}$ to $B = 1.5 \times 2 \times B_\mathrm{Auger} = 3 \times B_\mathrm{Auger}$. 

For the same reason, this saturation effect sets in later for air showers with smaller values of $\sin \upalpha$ as the Lorentz force is proportional to $\sin \upalpha \times B_\mathrm{Earth}$.
Hence, at locations with strong geomagnetic field strengths also a deviation from a quadratic scaling with $\sin \upalpha$ is expected. For $B = B_\mathrm{Auger}$ we do not observe any deviation from a quadratic scaling with $\sin \upalpha$.
In addition, we observe that the saturation effect is stronger pronounced for larger zenith angles where the air shower develops in a part of the atmosphere with smaller density and traverses more geometrical distance during the shower development. Hence, the shower particles are longer, and therefore also stronger, accelerated by the geomagnetic field which results in an earlier onset of saturation effects. 
All this higher order dependencies support the explanation presented above where a saturation effect is expected.

A recent analysis of the impact of thunderstorm electric fields on the radio emission strength \cite{TrinhScholtenBuitinkEtAl2016} also reported that the radio emission does not scale proportional to the atmospheric electric field component that is oriented into the direction of the Lorentz force. At larger field strengths they also observed a saturation effect that was explained by the change in coherence length scales. Due to the larger transverse velocity, electrons and positrons trail further than the coherence length behind the shower front. This behavior might also or partly explain the saturation effects that we observe.

However, within the variations of the geomagnetic field strength occurring on Earth and the desired accuracy of this analysis, this second order dependence can be neglected. 
Hence, we extend Eq.~\eqref{eq:Srddxmax} by the geomagnetic field dependence to 
\begin{equation}
S_\mathrm{RD}^{\rho} = \frac{E_\mathrm{rad}}{a'(\rho_\mathrm{X_\mathrm{max}})^2 + (1-a'(\rho_\mathrm{X_\mathrm{max}})^2) \sin^2\upalpha \left(\frac{B_\mathrm{Earth}}{\unit[0.243]{G}}\right)^{1.8}} \,\, \frac{1}{\left(
1 - p_0 + p_0 \,  \exp[p_1 (\rho_{X_\mathrm{max}} - \langle \rho \rangle)]\right)^2 } \,
 \label{eq:SrddxmaxB}
\end{equation}
with
\begin{equation}
 a'(\rho_\mathrm{X_\mathrm{max}}) = a(\rho_\mathrm{X_\mathrm{max}}) / (B_\mathrm{Earth}/\unit[0.243]{G})^{0.9} \,
\end{equation}
as the relative strength of the charge-excess component $a$ also changes with changing geomagnetic field.
This formula is then applicable to any location on Earth.

\subsection{Suitable Energy Estimator for a Measurement}

In a measurement, \xmax is not always accessible or may have large experimental uncertainties. As most of the variation in \rhoxmax is due to the zenith angle of the shower direction, we implement a correction that takes only the zenith angle into account. For all air showers we assume the same average $\langle X_\mathrm{max} \rangle$ of \unit[669]{g/cm$^2$} to calculate the density at the position of the shower maximum $\rho_\theta = \rho(\theta, \langle X_\mathrm{max} \rangle)$. We first parametrize the variation of the charge-excess fraction using the zenith angle and find  
\begin{equation}
 a(\rho_\theta) = 0.45  \left(\exp[\unit[1.14]{m^3/kg} \, (\rho_\theta- \langle \rho \rangle)]\right) - 0.24\, ,
 \label{eq:azenith}
\end{equation}
and then define the energy estimator as
\begin{equation}
S_\mathrm{RD}^{\rho_\theta} = \frac{E_\mathrm{rad}}{a'(\rho_\theta)^2 + (1-a'(\rho_\theta)^2) \sin^2\upalpha \left(\frac{B_\mathrm{Earth}}{\unit[0.243]{G}}\right)^{1.8}} \,\, \frac{1}{\left(
1 - p_0 + p_0 \,  \exp[p_1 (\rho_\theta - \langle \rho \rangle)]\right)^2 } \,
 \label{eq:Srdzenith}
\end{equation}
with 
\begin{equation}
 a'(\rho_\theta) = a(\rho_\theta) / (B_\mathrm{Earth}/\unit[0.243]{G})^{0.9} \, .
\end{equation}

Again, we perform a chi-square fit and find the parameters listed in Tab.~\ref{tab:srdzenith}. We determine the energy resolution by fitting a Gaussian function to the scatter around the calibration curve and find a standard deviation of 4\%.
\begin{table}[htb]
\centering
\begin{tabular}{cr@{ $\pm$ }l}
 \hline \hline
 $A$ & 1.629 & 0.003 \\
 $B$ & 1.980 & 0.001 \\
 $p_0$ & 0.239 & 0.007 \\
 $p_1$ & -3.13 & 0.07 \unit{m$^3$/kg}\\
 \hline
 \hline
\end{tabular}
\caption{Best-fit parameters for an energy estimator with zenith angle correction as defined in Eq.~\eqref{eq:Srdzenith}.}
\label{tab:srdzenith}
\end{table}

The slope parameter $B$ shows a deviation from 2 that is significantly larger than the uncertainties. This is because the average \xmax increases linearly with the logarithm of the cosmic-ray energy \cite{AugerXmaxICRC2015}. Hence, for the same zenith angle, more energetic showers develop deeper in the atmosphere, i.e., in a region of higher atmospheric density, and consequently have a smaller radiation energy. As we do not correct for an \xmax dependence, the effect appears with a value slightly smaller than 2 in the slope parameter $B$. However, the energy resolution achieved using only a zenith angle correction is still only 4\%, which is small compared to current experimental uncertainties \cite{AERAEnergyPRD}.

\subsection{Clipping}
If the radiation energy is detected by an observer at a height $h$ above sea level, the air shower may not have released all its radiation energy. The shower might not be fully developed when it reaches the ground, i.e., some part of the shower that would have contributed to the radiation energy is clipped. 
The strength of the clipping depends on the atmospheric depth between the observer and the shower maximum \xmax and is defined as
\begin{equation}
D_\mathrm{X_\mathrm{max}} = \int\limits_{h_0}^\infty \frac{\rho(h)}{\cos \theta} \, \mathrm{d}h - X_\mathrm{max} = \int\limits_{h_0}^{h(X_\mathrm{max})} \frac{\rho(h)}{\cos \theta} \, dh  ,
\label{eq:dxmax}
\end{equation}
where $h_0$ is the altitude of the observer, $\theta$ is the zenith angle and $\rho(h)$ the atmospheric density at height $h$ above sea level. 

\begin{figure}[bt]
 \centering
 \includegraphics[width=0.49\textwidth]{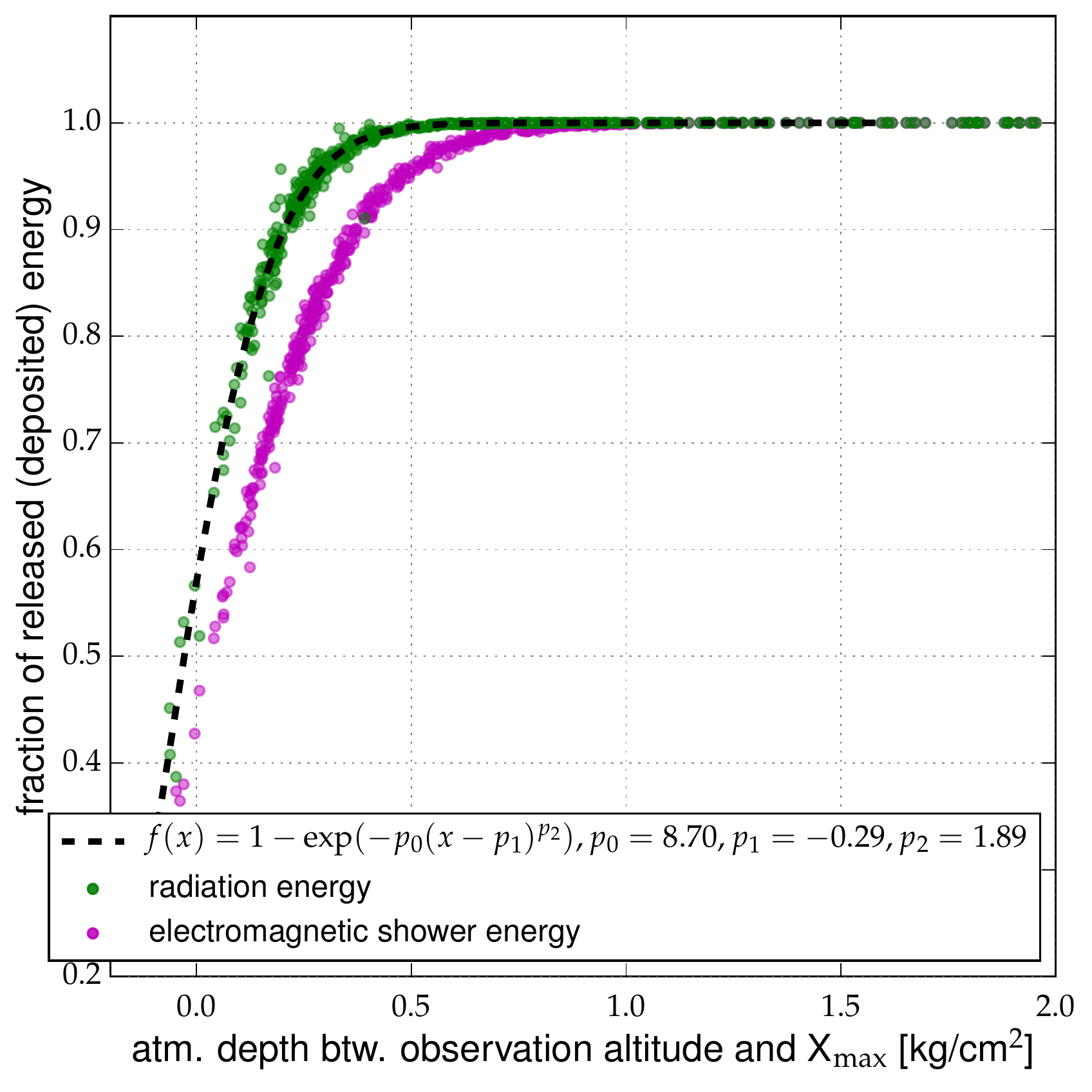}
 \includegraphics[width=0.49\textwidth]{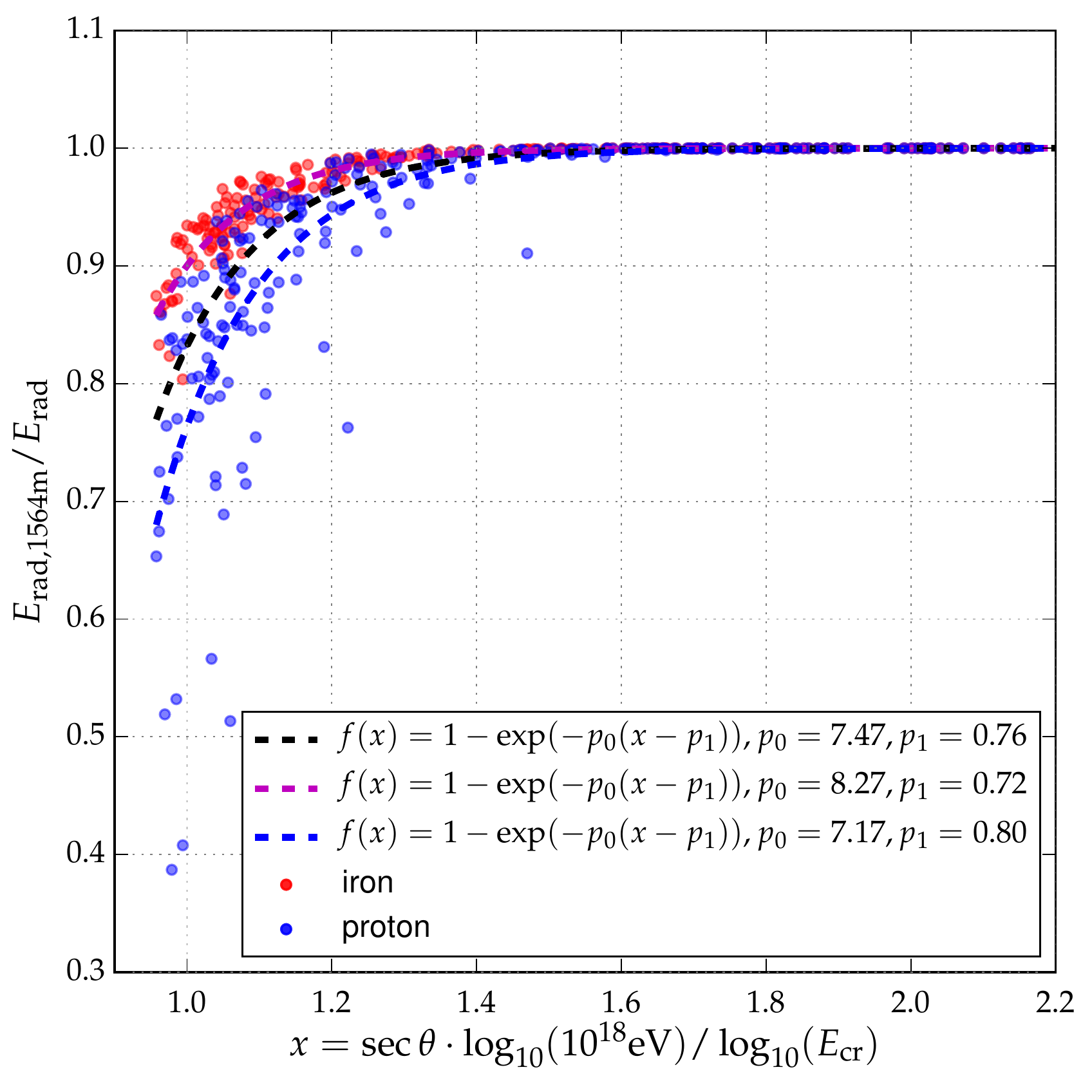}
 \caption{(left) The green circles show the fraction of radiation energy that is released in the atmosphere as a function of the atmospheric depth between the observation height and shower maximum. The dashed line shows a parametrization of this dependence. The magenta circles show the fraction of the deposited energy of the electromagnetic part of the air shower in the atmosphere. (right) Fraction of radiation energy that is released until a height of \unit[1564]{m a.s.l.} as a function of the zenith angle and cosmic-ray energy. Red circles represent iron-induced air showers and blue circles show proton-induced air showers. The dashed lines show parametrizations of the dependence for all data points (black), proton showers (blue) and iron showers (magenta).}
 \label{fig:clipping}
\end{figure}

Fig.~\ref{fig:clipping} shows the fraction of radiation energy released in the atmosphere as a function of \dxmax. This dependence is independent of observation altitude, as we directly calculate the atmospheric depth between the observation altitude and \xmax. An observer at higher altitude observes smaller ``distances to \xmax'' and is influenced more strongly by clipping. We parametrize the dependence using the following function
\begin{equation}
 \frac{E_\mathrm{rad}(\dxmaxm)}{E_\mathrm{rad}} = 1- \exp \left(\unit[-8.7]{cm^2/kg} \,  (\dxmaxm + \unit[0.29]{kg/cm^2})^{1.89}\right) \, .
 \label{eq:clippingdxmax}
\end{equation}
The clipping of the electromagnetic part of the air shower is significantly stronger, which is also shown in Fig.~\ref{fig:clipping} on the left. This is in agreement with the previous finding that the longitudinal profile of the radiation energy release is shifted towards smaller atmospheric depths compared to the energy deposit of the electromagnetic part of the air shower (cf. Fig.~\ref{fig:Long1}). At the point where only 75\% of the electromagnetic part of the air shower has deposited its energy in the atmosphere, 90\% of the radiation energy has already been released.

This effect is of prime importance in an experiment, as the radiation energy needs to be corrected for clipping to correctly estimate the cosmic-ray energy. As \xmax is not always accessible or may have large experimental uncertainties, we present an alternative parametrization of the clipping effect for different observation heights using the zenith angle and the cosmic-ray energy. The atmospheric depth between the observer and shower maximum is proportional to the secant of the zenith angle with a second-order dependence on the cosmic-ray energy as the average \xmax increases proportionally to the logarithm of the energy (cf. Eq.~\eqref{eq:dxmax}). Fig.~\ref{fig:clipping} on the right shows this parametrization for an observation height of \unit[1564]{m a.s.l.}, which corresponds to the altitude of the radio detector of the Pierre Auger Observatory \cite{ICRC2015JSchulz}. A significant difference between air showers induced by protons or iron nuclei is visible, as iron showers have a smaller average \xmax than proton showers. We again parametrize the dependence with an exponential function and provide the best-fit values for a mixed, a pure proton and a pure iron composition in the legend of the figure and in Tab.~\ref{tab:clipping}.

\begin{table}[tb]
\centering
\begin{tabular}{lcc}
 \hline \hline
 \textbf{composition} & \textbf{$\boldsymbol{p_0}$} & \textbf{$\boldsymbol{p_1}$} \\ \hline
 \hspace*{0.55cm}\textbf{\unit[1564]{m a.s.l.}} (Auger) & & \\
 \hspace*{0.55cm}\hspace*{0.55cm}pure proton & 7.17 & 0.8 \\
 \hspace*{0.55cm}\hspace*{0.55cm}pure iron & 8.27 & 0.72 \\
 \hspace*{0.55cm}\hspace*{0.55cm}50\% proton, 50\% iron & 7.46 & 0.76 \\
 
 \hspace*{0.55cm}\textbf{\unit[675]{m a.s.l.}} (Tunka-Rex) & & \\
 \hspace*{0.55cm}\hspace*{0.55cm}pure proton & 9.57 & 0.77 \\
 \hspace*{0.55cm}\hspace*{0.55cm}pure iron & 10.22 & 0.68 \\
 \hspace*{0.55cm}\hspace*{0.55cm}50\% proton, 50\% iron & 9.73 & 0.74 \\
 
 \hspace*{0.55cm}\textbf{\unit[0]{m a.s.l.}} (LOFAR) & & \\
 \hspace*{0.55cm}\hspace*{0.55cm}pure proton & 11.93 & 0.76 \\
 \hspace*{0.55cm}\hspace*{0.55cm}pure iron & 11.85 & 0.66 \\
 \hspace*{0.55cm}\hspace*{0.55cm}50\% proton, 50\% iron & 11.9 & 0.72 \\
 \hline
 \hline
\end{tabular}
\caption{Parametrization of the clipping effect using the zenith angle $\theta$ and the cosmic ray energy $E_\mathrm{cr}$ for different heights above sea level with the function $f(x) = 1 - \exp(-p_0(x-p_1))$ with $x = \sec \theta \cdot \log_{10}(10^{18}~\mathrm{eV})/\log_{10}(E_\mathrm{cr})$.}
\label{tab:clipping}
\end{table}

\begin{figure}[bt]
 \centering
 \includegraphics[width=0.48\textwidth]{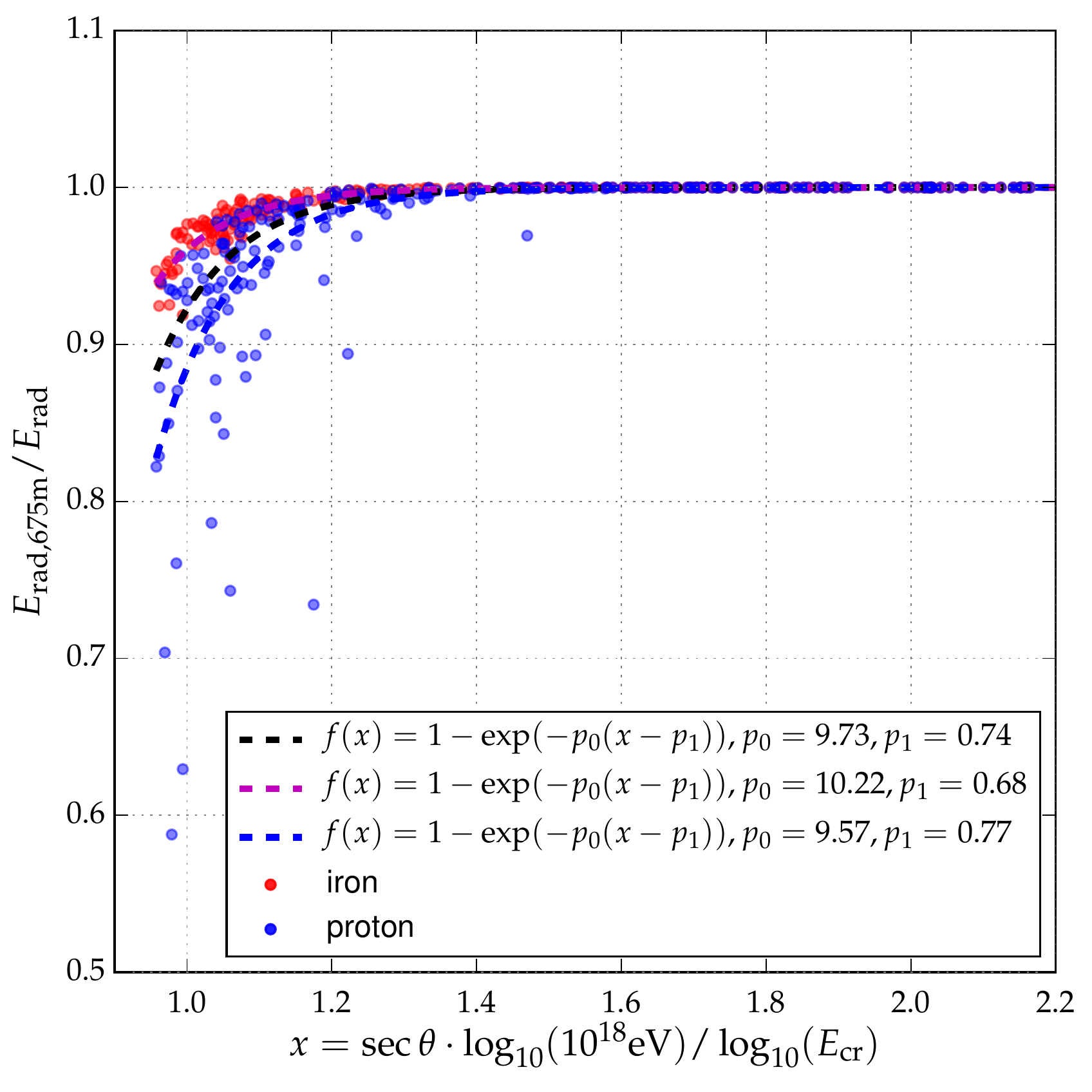}
 \includegraphics[width=0.48\textwidth]{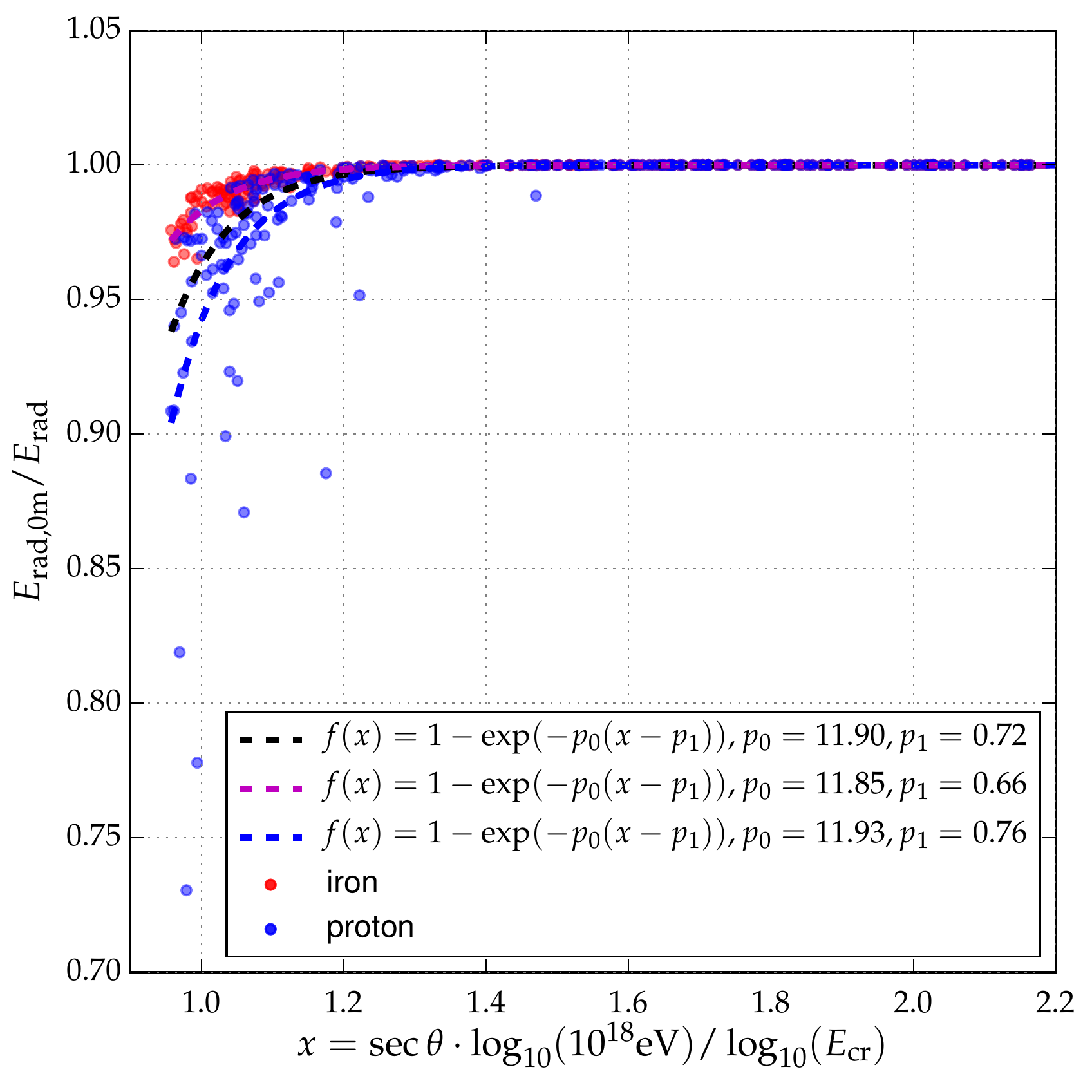}
 \caption{Fraction of the radiation energy that is generated up to a height of \unit[675]{m a.s.l.} (left) and \unit[0]{m a.s.l.} (right) as a function of the zenith angle and cosmic-ray energy. Red circles represent iron-induced air showers and blue circles show proton-induced air showers. The dashed lines show parameterizations of the dependence for all data points (black), proton showers (blue) and iron showers (magenta).}
 \label{fig:clipping_sealeveltunka}
\end{figure}

We repeated the analysis for an observation height of \unit[675]{m asl} which corresponds to the location of the Tunka-Rex experiment \cite{TunkaRex2015} and for an observation height of \unit[0]{m a.s.l.} which corresponds to the location of the LOFAR experiment \cite{LOFARICRC2015} in Fig.~\ref{fig:clipping_sealeveltunka} and Tab.~\ref{tab:clipping}.

Alternatively, the parametrization of Eq.~\eqref{eq:clippingdxmax} together with the parametrization of the average \xmax as a function of cosmic-ray energy and mass using different interaction models \cite{Domenico2013} can be used to calculate the clipping effect for any observation altitude and assumption of the average \xmax.

\section{Influence of Settings of the Air-Shower Simulation}
In this section, we study the influence of different settings of the air-shower simulation on the radiation energy in the \unit[30-80]{MHz} frequency band. In each cross-check we follow the same procedure. For the different settings of the simulation, we simulate several air showers with fixed geometry and energy. The incoming direction of the air shower is fixed to 50$^\circ$ zenith angle coming from the south and we use only iron primaries to minimize \xmax fluctuations. We simulate an observer at sea level so that we can exclude any clipping effect for this geometry. Unless otherwise stated, the same simulation setup as described in Sec.~\ref{sec:simulation_setup} is used.

We correct the resulting radiation energy for its $\sin \upalpha$ and \xmax dependence according to Eq.~\eqref{eq:Srddxmax} and normalize $S_\mathrm{RD}^\rho$ to the mean electromagnetic shower energy of the set of simulated air showers by multiplying $S_\mathrm{RD}^\rho$ with $(\langle E_\mathrm{em} \rangle / E_\mathrm{em})^{2}$, as the amount of cosmic-ray energy that ends up in the electromagnetic shower energy is slightly different for each shower. Thereby, we correct for all known dependencies and are most sensitive to the influence of different settings in the simulation. However, the effect of different settings is mostly smaller than the random fluctuations in the radiation energy. Hence, we simulate at least 20 air showers for each setting of the simulation and compare the mean of the normalized radiation energies.

\subsection{Impact of Hadronic Interaction Models}

We investigated the impact of the choice of the high-energy hadronic interaction model on the radiation energy by simulating 39 air showers for each of the two post-LHC models QGSJeII-04 \cite{QGSJet} and EPOS-LHC \cite{EPOSLHC}. We use the settings described above and fix the cosmic-ray energy to \unit[1]{EeV}. For both interaction models we obtain consistent normalized radiation energies of \unit[10.97 $\pm$ 0.03]{MeV} (QGSJetII-04) and \unit[10.99 $\pm$ 0.02]{MeV} (EPOS-LHC). The quoted uncertainty is the uncertainty of the mean. We conclude that the effects of the high-energy hadronic interaction models are negligible for the prediction of the radiation energy.

We also checked for the influence of the low-energy hadronic interaction models and compared the two models FLUKA \cite{fluka} and UrQMD \cite{URQMD}. We used the same settings as above, allow QGSJetII-04 to treat the high-energy hadronic interactions and obtain a mean normalized radiation energy of \unit[10.97 $\pm$ 0.03]{MeV} using FLUKA and \unit[10.95 $\pm$ 0.04]{MeV} using UrQMD. We conclude that the choice of low-energy hadronic interaction model also has a negligible influence on the radiation energy.

\subsection{Impact of Approximations in the Air-Shower Simulation}
As a full simulation of all shower particles is not feasible at high cosmic-ray energies due to the large number of shower particles, only a representative sub-sample of particles is tracked.  This approach is known as ``thinning''. In this analysis we apply thinning at a level of $10^{-6}$ with optimized weight limitation \cite{Kobal2001}. To test the impact of the thinning level on radiation energy, we simulate air showers with thinning levels of $10^{-3}$, $10^{-4}$, $10^{-5}$, $10^{-6}$ and $10^{-7}$ and cosmic-ray energies of \unit[1]{EeV}, \unit[10]{EeV} and \unit[100]{EeV} and calculate the normalized radiation energy. In other words, we calculate the corrected radiation energy $S_\mathrm{RD}^\rho$ and for each set of showers with the same cosmic-ray energy we normalize $S_\mathrm{RD}^\rho$ to the average electromagnetic shower energy of this set. The results are presented in Tab.~\ref{tab:thinning}.

Regardless of the air-shower energy we obtain consistent results for thinning levels of $10^{-5}$, $10^{-6}$ and $10^{-7}$. For lower thinning levels we obtain larger radiation energies and observe an increased spread. On average, simulations with a thinning level of $10^{-4}$ produce $\sim$1\% more radiation energy and simulations with a thinning level of $10^{-3}$ produce $\sim$6\% more radiation energy. We conclude that thinning at a level of $10^{-6}$ does not introduce any bias in the simulation of the radiation energy.

The runtime of the simulation depends strongly on the thinning level. Relative to the runtime of a simulation with a thinning level of $10^{-6}$, a simulation with a thinning level of $10^{-3}$ ($10^{-4}$, $10^{-5}$) needs only 1\% (6\%, 20\%) of the runtime, whereas a simulation with a thinning level of $10^{-7}$ needs six times more computing time. Hence, even a stronger thinning at a level of $10^{-5}$ could be used in future to further speed up the simulations.  

\begin{table}[tb]
\centering
\renewcommand{\arraystretch}{1.1}
\begin{tabular}{cr@{~$\pm$~}l@{ }r@{~}lr@{~$\pm$~}l@{ }r@{~}lr@{~$\pm$~}l@{ }r@{~}l}
\hline \hline
& \multicolumn{12}{c}{\textbf{cosmic-ray energy}} \\
\textbf{thinning level} & \multicolumn{4}{c}{\unit[1]{EeV}}& \multicolumn{4}{c}{\unit[10]{EeV}}& \multicolumn{4}{c}{\unit[100]{EeV}} \\ \hline
$10^{-3}$&11.7&0.1&(1)&MeV&1285&14&(100)&MeV&139&2&(12)&GeV\\
$10^{-4}$&11.10&0.05&(0.34)&MeV&1234&6&(42)&MeV&131.7&0.5&(3.3)&GeV\\
$10^{-5}$&10.91&0.03&(0.22)&MeV&1214&4&(19)&MeV&129.7&0.5&(2.1)&GeV\\
$10^{-6}$&10.97&0.03&(0.21)&MeV&1216&3&(13)&MeV&129.8&0.4&(1.6)&GeV\\
$10^{-7}$&10.96&0.04&(0.19)&MeV&1216&4&(19)&MeV&129.8&0.3&(1.2)&GeV\\
\hline
\end{tabular}

\renewcommand{\arraystretch}{1}
\caption{The table shows the normalized radiation energies for different cosmic-ray energies and thinning levels. The geometry of the air showers is fixed to a zenith angle of 50$^\circ$ coming from the south. Each cell shows the mean of at least 20 air showers simulated with the same settings, but different random seeds. The uncertainties shown are the uncertainty of the mean, and the standard deviations are shown in brackets.}
\label{tab:thinning}
\end{table}

Another approximation in the simulation is that each shower particle is only followed until it reaches a minimum energy. The threshold energy is different for different particle species. In this analysis, we use a minimum energy of \unit[100]{MeV} for hadrons, \unit[50]{MeV} for muons and \unit[250]{keV} for the electromagnetic shower particles (electrons, positrons and photons). We check for the impact of the energy threshold of the electromagnetic particles that are relevant for the generation of radio emission by simulating iron-induced air showers with a fixed geometry and cosmic-ray energy, but different energy thresholds of \unit[250]{keV}, \unit[150]{keV} and \unit[80]{keV}. We use the same settings as in the previous cross-checks and simulate air showers with cosmic-ray energies of \unit[1]{EeV} and \unit[10]{EeV} and calculate the normalized radiation energy as in the thinning cross-check. The results are presented in Tab.~\ref{tab:ecuts}. We observe a slight trend towards smaller average radiation energies for lower energy thresholds. This is not a violation of energy conversation as in the endpoint formalism \cite{Ludwig2011438, Endpoint2011} of CoREAS the radiation from stopping the particle is already considered and simulating the movement of a particle up to small energies results in less radiation than an instantaneous stop. Also lowering the photon energy threshold leads to the consideration of low energetic Bremsstrahlungs photons resulting in a less straight movement of electrons and consequently less coherent radiation. However, for a \unit[1]{EeV} air shower, the shift in the average radiation energy is only 1\%, which would correspond to a shift of 0.5\% in the cosmic-ray energy. Hence, for the desired accuracy of the analysis this effect is negligible.

Again, the runtime depends strongly on the energy threshold. A simulation with a lower energy cut of \unit[150]{keV} (\unit[80]{keV}) runs 1.4 times (3 times) longer than a simulation with the default energy cut of \unit[250]{keV}.

\begin{table}[tb]
\centering
\renewcommand{\arraystretch}{1.1}
\begin{tabular}{ccc}
\hline \hline
& \multicolumn{2}{c}{\textbf{cosmic-ray energy}} \\
\textbf{energy cut} & \unit[1]{EeV}& \unit[10]{EeV} \\ \hline
\unit[250]{keV}&\unit[10.97 $\pm$ 0.03 (0.21)]{MeV}&\unit[1217 $\pm$ 3 (13)]{MeV}\\
\unit[150]{keV}&\unit[10.91 $\pm$ 0.04 (0.19)]{MeV}&\unit[1205 $\pm$ 3 (14)]{MeV}\\
\unit[80]{keV}&\unit[10.84 $\pm$ 0.03 (0.15)]{MeV}&\unit[1197 $\pm$ 3 (12)]{MeV}\\
\hline
\end{tabular}
\renewcommand{\arraystretch}{1}
\caption{The table shows the normalized radiation energies for different cosmic-ray energies and lower energy cuts for the electromagnetic shower particles. The geometry of the air showers is fixed to a zenith angle of 50$^\circ$ coming from south. Each cell shows the mean of at least 20 air showers simulated with the same settings but different random seeds. The uncertainties shown are the uncertainty of the mean, and the standard deviations are shown in brackets. The percentage values denote the relative deviations to the reference refractivity.}
\label{tab:ecuts}
\end{table}

\subsection{Influence of Air Refractivity}
As already visible from Fig.~\ref{fig:LDFexample}, the refractive index of the atmosphere has an influence on radiation energy. In the simulation, we specify the refractivity at sea level which is then scaled proportionally to higher altitudes with the air density. To estimate the influence of the air refractivity on radiation energy we vary the value of $n-1$. We again simulate at least 20 air showers induced by iron nuclei for each choice of refractivity for a cosmic-ray energy of \unit[1]{EeV} and a fixed geometry as in the previous cross-checks and calculate the normalized radiation energy. The results are presented in Tab.~\ref{tab:refractivity}. A change in the refractivity of +10\% (-10\%) results in a 3\% larger (3\% smaller) radiation energy and a change of $\pm$5\% in refractivity results in a shift of $\pm$1.5\% in radiation energy. We note again that the $\pm$3\% ($\pm$1.5\%) shift in radiation energy would correspond to a $\pm$1.5\% ($\pm$0.8\%) shift in the electromagnetic shower energy, as the radiation energy scales quadratically with the electromagnetic shower energy. 

We determine the variations in refractivity using data from the Global Data Assimilation System (GDAS). This system combines several meteorological measurements with numerical weather predictions and provides the main state variables of the atmosphere on a $1^\circ\times1^\circ$ latitude longitude grid every three hours. Among other things the temperature, pressure and humidity at the surface is provided from which we calculate the refractivity. The yearly fluctuations of the air refractivity at the site of the Pierre Auger Observatory are 7\%, and are only 4\% and 3\% at the LOFAR and Tunka sites, respectively. 
Hence, this environmental dependence is small compared to current experimental uncertainties. 

We also calculated the average refractivity at sea level by rescaling the average refractivity at the surface to its value at sea level using the model of the US standard atmosphere. We find a refractivity at sea level of $3.1 \times 10^{-4}$ for the Pierre Auger and Tunka site and $3.2 \times 10^{-4}$ for the LOFAR site.

\begin{table}[tb]
\centering
\renewcommand{\arraystretch}{1.1}
\begin{tabular}{r@{ }rcr@{ }r}
\hline \hline
\multicolumn{3}{c}{\textbf{refractivity at sea level}} & \multicolumn{2}{c}{\textbf{normalized radiation energy}} \\ \hline
\noalign{\smallskip}
$2.04 \times 10^{-4}$&(-30\%)& &\unit[9.84 $\pm$ 0.04 (0.17)]{MeV} &(-10.3\%)\\
$2.63 \times 10^{-4}$&(-10\%)& &\unit[10.63 $\pm$ 0.04 (0.18)]{MeV} &(-3.1\%)\\
$2.77 \times 10^{-4}$&(-5\%)& &\unit[10.79 $\pm$ 0.05 (0.20)]{MeV} &(-1.6\%)\\
$2.92 \times 10^{-4}$& &&\unit[10.97 $\pm$ 0.03 (0.21)]{MeV} & \\
$3.07 \times 10^{-4}$&(+5\%)& &\unit[11.12 $\pm$ 0.05 (0.21)]{MeV} &(+1.4\%)\\
$3.21 \times 10^{-4}$&(+10\%)&&\unit[11.29 $\pm$ 0.04 (0.19)]{MeV} &(+2.9\%)\\
$3.80 \times 10^{-4}$&(+30\%)&&\unit[11.82 $\pm$ 0.05 (0.21)]{MeV} &(+7.7\%)\\
\hline
\end{tabular}
\renewcommand{\arraystretch}{1}
\caption{The table shows the normalized radiation energies for different refractivities of the atmosphere. The geometry of the air showers is fixed to a zenith angle of 50$^\circ$ coming from south. Each cell shows the mean of at least 20 air showers simulated with the same settings but different random seeds. The uncertainties shown are the uncertainty of the mean, and the standard deviation is shown in brackets.}
\label{tab:refractivity}
\end{table}

\section{Conclusion}

We studied the energy released in air showers in the form of MHz radiation using CoREAS air-shower simulations. The radiation energy is determined by the integration over the energy fluence in the shower plane. We present an efficient method that exploits the knowledge of the emission mechanisms and requires a simulation of the energy fluence at only one axis in the shower plane. Using this information and the method developed in this contribution, the energy fluence can be calculated at any position in the shower plane. 

Depending on the distance between the observer and the region in the atmosphere where the radiation is released, the shape of the signal distribution on the ground changes significantly. For small distances to the emission region, the signal distribution is narrow around the shower axis with large energies per unit area, whereas for large distances to the emission region the radiation energy is distributed over a larger area resulting in a broad signal distribution with a small amount of energy per unit area. 
As soon as the air shower has emitted all its radiation energy, the total radiation energy, i.e., the integral over the signal distribution on the ground, remains constant. In particular, it does not depend on the signal distribution on the ground or on the observation altitude and is thus directly comparable between different experiments.

We studied the longitudinal profile of the radiation energy release, i.e., how much radiation energy is released at what atmospheric depth, and found that it can be described well by a Gaisser-Hillas function with a maximum before the maximum of the longitudinal profile of the energy deposit by the electromagnetic part of the air shower where the average offset is \unit[48]{g/cm$^2$}.

Using our method we can determine the radiation energies originating from the geomagnetic and charge-excess emission process separately. We find that the ratio of the two radiation energies is not constant but depends on the air density where the radiation is generated. 

The radiation energy is well suited to determine the cosmic-ray energy. The radiation energy -- corrected for the dependence on the geomagnetic field -- correlates best with the energy in the electromagnetic part of the air shower and exhibits quadratic scaling with the electromagnetic shower energy, as is expected for coherent emission. The electromagnetic shower energy can be converted to the primary cosmic-ray energy using predictions from hadronic interaction models or a direct measurement of the invisible energy fraction.

The radiation energy has a second-order dependence on the air density of the emission region. After correcting this effect, the corrected radiation energy and the electromagnetic shower energy have a scatter of less than 3\%.

In addition, we present a more practical parametrization of the dependence between radiation energy and electromagnetic shower energy using only the geometry of the air shower, i.e., without using \xmax information, and obtain a resolution of 4\%. This scatter of 4\% is well below current experimental uncertainties, so that the radiation energy is well suited to estimate the cosmic-ray energy.

If the radiation energy is detected at a particular observation height, the air shower may not have released all its radiation energy. The strength of this effect depends on the atmospheric depth between observation height and shower maximum. We present a parametrization of this effect that can be used in experiments to correctly determine the full radiation energy and thereby estimate the cosmic-ray energy. The radiation energy is influenced less by clipping than the electromagnetic part of the air shower as the radiation energy is released earlier in the shower development.

We studied the effects of specific settings of the air-shower simulation on the radiation energy. We found that the choice of hadronic interaction model has a negligible influence on the radiation energy as well as our choice of thinning level and lower energy thresholds in the simulation. A variation of the air refractivity in a realistic range of $\pm$5\% results in a variation of the radiation energy of merely $\pm$1.5\%. 

We conclude that the radiation energy released by extensive air showers is well understood and can be used to precisely determine the cosmic-ray energy as well as to cross-calibrate different experiments.

\acknowledgments
We gratefully acknowledge discussions within the AERA radio working group of
the Pierre Auger Collaboration, and the assistance of Florian Gate in
processing the GDAS profiles. We wish to thank Olaf Scholten and Frank
Schröder for helpful comments on the manuscript including the interpretation of the decrease of the charge-excess radiation energy release. We also thank Tanguy
Pierog and Dieter Heck for their support related to the CORSIKA software
and Gevorg Poghosyan and the IT Center of RWTH Aachen University for
their assistance with the MPI parallelization of CORSIKA.
This work is supported by the Deutsche Forschungsgemeinschaft, the
Ministerium f\"ur Wissenschaft und Forschung, Nordrhein-Westfalen, the
Bundesministerium f\"ur Bildung und Forschung (BMBF), and the Helmholtz
Alliance for Astroparticle Physics.

\appendix

\section{Approximations in the Calculation of the Radiation Energy}

\subsection{Approximation of the same phase between $\mathbf{\Egeo}$ and $\mathbf{\Ece}$}
\label{sec:approx1}
\begin{figure}[bt]
 \includegraphics[width=\textwidth]{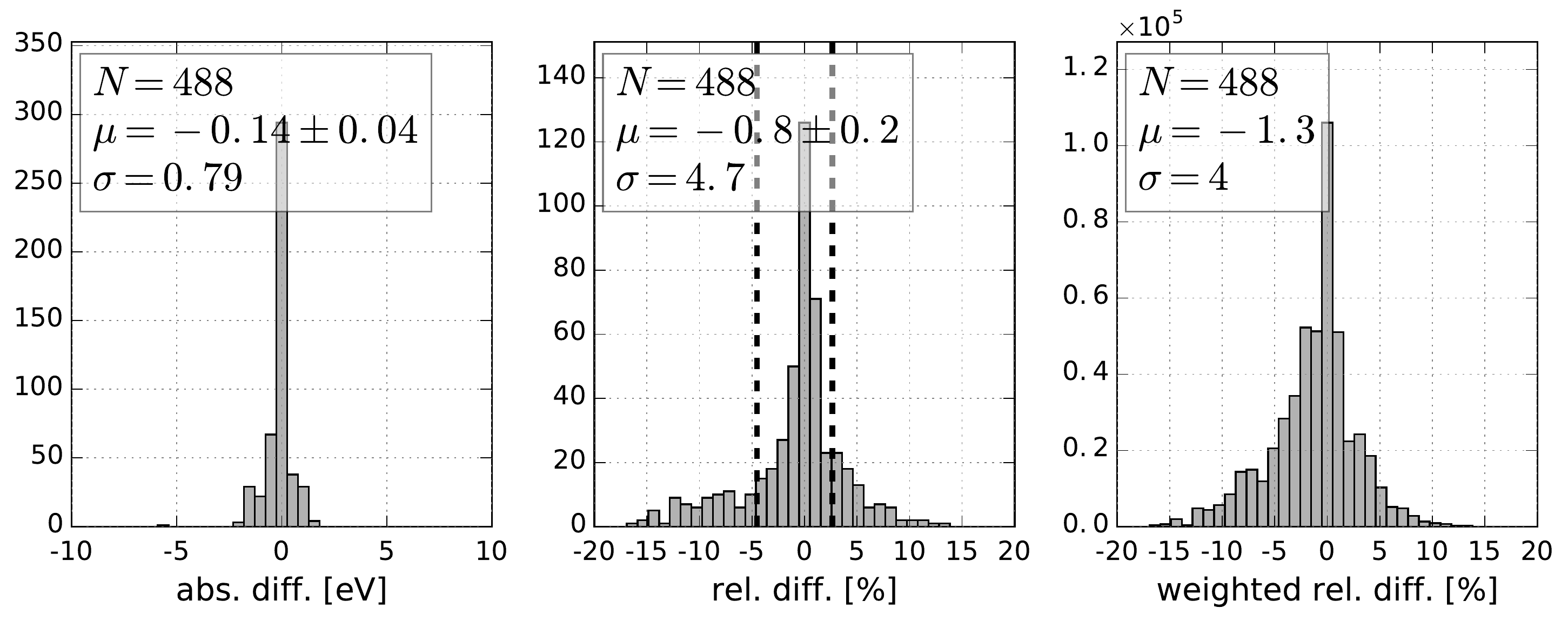}
 \caption{Difference between $\uvB$ calculated directly from $\EvB$ and $\uvB$ calculated via Eq.~\eqref{eq:uvB}. The dashed lines show the 68\% quantiles. The weight in the right histogram is $r \cdot \uvB$ with the distance $r$, i.e., it is weighted according to its influence on radiation energy.}
 \label{fig:uvB_approx}
\end{figure}

The validity of Eq.~\eqref{eq:uvB}, which is based on the assumption that radio pulses originating from the geomagnetic and charge-excess components are in-phase, can be validated by inspecting simulated radio pulses at the $\phi$ = 45$^\circ$, 135$^\circ$, 225$^\circ$, 315$^\circ$ directions. At these positions the charge-excess signal splits up equally into the $\EvB$ and $\EvvB$ polarizations. Hence, we can determine $\Ece(t)$ from $\EvvB(t)$ and $\Egeo(t)$ by subtracting $\Ece(t)$ from $\EvB(t)$:
\begin{eqnarray}
 \Ece(t) = &\sin^{-1} \phi \, \EvvB(t) \\
 \Egeo(t) = &\EvB(t) - \cos \phi \, \Ece(t) \stackrel{\phi = 45^\circ + n \cdot 90^\circ}{=} \EvB(t) - \EvvB(t) \, .
 \label{eq:geocedecomposition}
\end{eqnarray}
Then, we calculate $\uvB$ directly from $\EvB(t)$ and via Eq.~\eqref{eq:uvB} from $\Egeo(t)$ and $\Ece(t)$. Fig.~\ref{fig:uvB_approx} shows the difference for eight different simulated air showers (E = \unit[0.5]{EeV}, zenith angle = 30$^\circ$, 40$^\circ$, 50$^\circ$, 60$^\circ$, azimuth angle = 180$^\circ$, 270$^\circ$). The distributions are centered around zero, the mean is slightly ($\sim$1\%) shifted to negative differences, meaning that the approximation using Eq.~\eqref{eq:uvB} slightly overestimates the true energy density. 

\subsection{Approximation of radial symmetry of geomagnetic and charge-excess LDFs}
\label{sec:approx2}
\begin{figure}[bt]
\centering
 \includegraphics[width=1\textwidth]{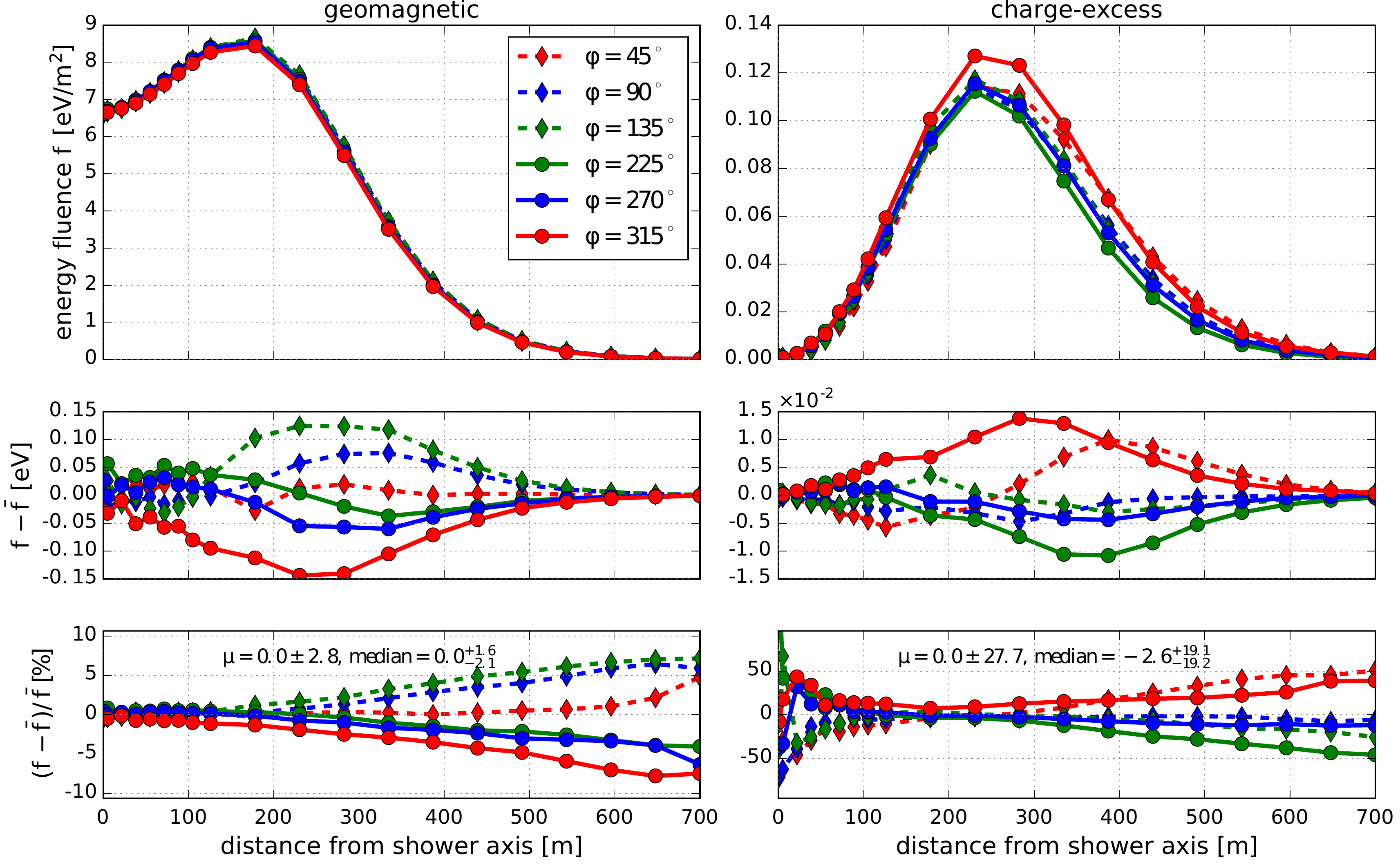}
 \caption{Geomagnetic and charge-excess lateral signal distribution for six different azimuthal directions of an \unit[0.5]{EeV} iron-induced air shower coming from the south with a zenith angle of 60$^\circ$. The lower plots show the absolute and relative deviation to the average LDF of the six azimuthal directions.}
 \label{fig:radsymmetry}
\end{figure}

The decomposition of the signal into the geomagnetic and charge-excess contribution of Eq.~\eqref{eq:geocedecomposition} can be used to check for the radial symmetry of the geomagnetic and charge-excess LDFs. Fig.~\ref{fig:radsymmetry} shows the energy density versus the distance to the shower axis for the geomagnetic and charge-excess part for different azimuthal directions $\phi$ (cf. Fig.~\ref{fig:starpattern}).  The geomagnetic LDF exhibits only slight deviations from a radially symmetric LDF. The 68\% quantile of the relative deviation is below 2\%. 

The relative deviations of the charge-excess LDF for different azimuthal directions are larger than for the geomagnetic LDF. The 68\% quantiles vary between 12\% and 20\%. However, on an absolute scale the deviations are smaller than for the geomagnetic LDF. This could mean that the observed deviations, or at least a large part of them, are due to random fluctuations in the simulations. Even if there were a deviation from a radially symmetric charge-excess LDF it would have negligible influence on the total radiation energy. Only if the radiation energy of the charge-excess component is investigated on its own, could this introduce additional uncertainty. 

\providecommand{\href}[2]{#2}\begingroup\raggedright\endgroup

\bibliographystyle{JHEP}

\end{document}